\documentclass[journal]{IEEEtran}
\usepackage{graphicx}

\usepackage{caption}
\usepackage[caption=false,font=footnotesize]{subfig} 

\usepackage{amsthm}
\usepackage{cite}

\usepackage[utf8]{inputenc}

\ifCLASSINFOpdf
\else
\fi
\usepackage{amsmath}
\usepackage{xcolor}

\usepackage{algorithmic}
\usepackage{algorithm}

\usepackage{bm}
\usepackage{graphicx}
\usepackage{array}
\usepackage{fixltx2e}

\usepackage{amssymb}

\begin{document}
%

\title{Robust Beamforming and Time Allocation for Time-Division Cell-Free Near-Field ISAC}
%
%
%

\author{Chaedam~Son, Si-Hyeon Lee~\IEEEmembership{Senior Member,~IEEE}
\thanks{} 
\thanks{Chaedam Son and Si-Hyeon Lee are with the School of Electrical Engineering of Korea Advanced Institute of Science and Technology (KAIST), Daejeon 34141, South Korea. (e-mail: scd5929@kaist.ac.kr and sihyeon@kaist.ac.kr). This work has been submitted to the IEEE for possible publication. Copyright may be transferred without notice, after which this version may no longer be accessible.}
}

%
%

\setlength{\arraycolsep}{5pt}    
\renewcommand{\arraystretch}{0.9}
%



\newtheorem{theorem}{Theorem}
\maketitle

\begin{abstract}
In this paper, we propose a time-division near-field integrated sensing and communication (ISAC) framework for cell-free multiple-input multiple-output (MIMO), where sensing and downlink communication are separated in time. During the sensing phase, user locations are estimated and used to construct location-aware channels, which are then exploited in the subsequent communication phase. By explicitly modeling the coupling between sensing-induced localization errors and channel-estimation errors, we capture the tradeoff between sensing accuracy and communication throughput. Based on this model, we jointly optimize the time-allocation ratio, sensing covariance matrix, and robust downlink beamforming under imperfect channel state information (CSI).
The resulting non-convex problem is addressed via a semidefinite programming (SDP)-based reformulation within an alternating-optimization framework. To further reduce computational complexity, we also propose two low-complexity suboptimal designs: an error-ignorant scheme and a maximum ratio transmission (MRT)-based scheme. Simulation results show that the proposed scheme significantly improves localization accuracy over far-field and monostatic setups, thereby reducing channel estimation errors and ultimately enhancing the achievable rate. {Moreover, the error-ignorant scheme performs well under stringent sensing requirements, whereas the MRT-based scheme remains robust over a wide range of sensing requirements by adapting the time-allocation ratio, albeit with some beamforming loss.}
\end{abstract}


%
\IEEEpeerreviewmaketitle

\section{Introduction}
\IEEEPARstart{I}{ntegrated} sensing and communication (ISAC) has emerged as a key enabling technology for sixth-generation (6G) wireless networks~\cite{ISAC_survey1,ISAC_survey2}. By allowing sensing and communication functionalities to operate simultaneously within a shared spectral and hardware framework, ISAC significantly enhances spectral efficiency and overall resource utilization. Moreover, accurate environmental sensing facilitates improved channel estimation and reduced pilot overhead, thereby enhancing communication reliability~\cite{ISAC_survey4}. 
In this context, the integration of ISAC with multiple-input multiple-output (MIMO) systems is particularly compelling. Large-scale antenna arrays provide high spatial resolution for sensing and enable sensing-assisted channel estimation, effectively addressing one of the fundamental bottlenecks in MIMO communications. {These complementary advantages have motivated extensive research on ISAC-enabled MIMO systems for reliable quality of service (QoS) under challenging propagation conditions~\cite{MIMO_ISAC_FULLduple}.}

However, as the number of antennas increases—resulting in larger array apertures—and communication frequencies rise, near-field effects have recently attracted increasing attention \cite{nearfield_isac_4(confer), nearfield_isac_5, nearfield_isac_6}. In the near-field regime, spherical wave propagation makes each array response depend on the user’s absolute position rather than only the AoA/AoD. This characteristic enables direct position inference from array responses given sufficient aperture, reducing reliance on distance-estimation techniques such as received signal strength indicator (RSSI), time difference of arrival (TDoA) or time of arrival (ToA). As a result, near-field ISAC has received increasing attention \cite{nearfield_isac_1, nearfield_isac_3(multi-target), nearfield_isac_2(velocity)}. In \cite{nearfield_isac_1}, the authors formulated an optimization problem to minimize the Cramér-Rao lower bound (CRLB) in a near-field single-target, multi-user scenario under user rate constraints. This work was later extended to a multi-target setting in~\cite{nearfield_isac_3(multi-target)}, focusing on minimizing energy while meeting both sensing and communication signal-to-interference-noise-ratio (SINR) requirements. Building on this line,~\cite{nearfield_isac_2(velocity)} incorporated velocity estimation and beamforming by exploiting the unique near-field property that each antenna observes distinct Doppler shifts due to spherical wavefronts. 

The performance of ISAC systems can be further enhanced by exploiting spatial diversity through multi-static sensing, where multiple spatially separated nodes collaboratively perform sensing and communication. In particular, cell-free architectures provide a natural and effective platform for realizing multi-static ISAC, as geographically distributed access points (APs) can jointly illuminate targets and collect sensing echoes from diverse spatial perspectives. Owing to these advantages, a large body of existing ISAC studies has adopted cell-free multi-static configurations, predominantly focusing on far-field sensing scenarios~\cite{cell-free1,cell-free2,cell-free3}. 
In \cite{cell-free1}, the authors defined the sensing signal-to-noise ratio (SNR) and the communication SINR in a cell-free setting and proposed methods to solve optimization problems whose objective is either communication or sensing performance. In \cite{cell-free2}, a cell-free scenario was considered where the distributed APs jointly serve the communication needs of users while sensing an eavesdropper.
Similarly, \cite{cell-free3} studied target detection algorithms in a cell-free system and optimized the corresponding power allocation. However, despite the extensive body of work on multi-static ISAC architectures in the far-field, their near-field counterparts have only begun to attract attention in very recent studies. In~\cite{nearfield_ISAC}, the authors investigated a near-field multi-static ISAC passive sensing system consisting of multiple communication users and a single sensing target. In this framework, multiple APs jointly transmit integrated sensing and communication signals, a designated AP collects the reflected echoes to detect and localize the target, and CSI is assumed to be available only in imperfect form.

In practice, however, acquiring CSI for all AP-user links in cell-free architectures is particularly challenging due to the large system dimensionality. In the near-field regime, channel estimation can be effectively supported by sensing, as the line-of-sight (LOS) component is typically dominant\cite{nearfield_isac_1,nearfield_isac_2(velocity),nearfield_isac_3(multi-target)} and the channel coefficients can be modeled as deterministic functions of user locations. Once the user positions are known, both the steering vectors and channel gains can be accurately reconstructed. This sensing-aided channel estimation is particularly well suited to multi-static scenario where sensing errors can be significantly reduced.
Motivated by these properties, we propose a time-division-based near-field ISAC framework for cell-free MIMO systems, in which user channels are estimated during a dedicated sensing phase and subsequently exploited in the communication phase. By decoupling sensing and communication and enabling location-aware channel construction, the proposed framework mitigates the channel-estimation burden---one of the major bottlenecks in near-field cell-free systems---while avoiding the high computational complexity inherent in fully joint sensing-and-communication designs.

{The main contributions of this paper are as follows:}

\begin{itemize}
  \item \textbf{ISAC framework:} 
We propose a time-division cell-free near-field ISAC system that differs from existing time-division ISAC approaches~\cite{time_allocation3,time_allocation4,time_allocation5,time_allocaion6}, which typically rely on fixed time-allocation ratios and perfect CSI assumptions. By explicitly modeling the coupling between sensing-induced localization errors and channel-estimation errors, we capture the fundamental tradeoff whereby allocating more time to sensing improves channel estimation accuracy at the expense of reduced communication time. By jointly optimizing the time-allocation ratio and robust downlink beamforming under channel uncertainty, we develop a practical and integrated ISAC framework that jointly accounts for sensing accuracy and communication throughput. 
  
  
  
  \item \textbf{Optimization algorithms:} 
  We optimize the sensing covariance matrix, downlink beamforming, and time-allocation ratio. 
  The sensing covariance matrix is first optimized independently. 
  Then, the downlink beamforming and time-division ratio are jointly optimized in an iterative \emph{alternating optimization (AO)} \cite{BCD} framework, 
  where the beamforming vectors are updated via a \emph{semidefinite programming (SDP)} \cite{convexopt} relaxation, and the time-division ratio is refined through constraint-margin maximization. 
  This joint design efficiently achieves robust convergence under imperfect CSI.

 ~Furthermore, to reduce computational complexity, we propose two suboptimal schemes: 
  (i) an \emph{error-ignorant} design that assumes perfect CSI for optimizing the beamforming and time-allocation ratio, and 
  (ii) an \emph{maximum ratio transmission (MRT)-based} scheme that leverages the near-field channel quasi-orthogonality property with large antenna arrays.
  
  \item \textbf{Numerical results:} 
  The simulation results demonstrate that the proposed near-field multi-static cell-free architecture achieves superior sensing performance compared to conventional single-cell and far-field systems. 
  {Furthermore, the proposed algorithms exhibit fast convergence and robustness to parameter variations.}
\end{itemize}

\textit{Notation}: Scalars, vectors, and matrices are denoted by regular, bold lowercase, and bold uppercase letters, respectively (e.g., $x$, $\pmb{x}$, $\pmb{X}$). The $(i,j)$-th element of a matrix $\pmb{X}$ is $[\pmb{X}]_{i,j}$, the $(p,q)$-th submatrix of a block matrix $\pmb{Y}$ is $[\pmb{Y}]_{\langle p,q\rangle}$, and the $i$-th element of a vector $\pmb{a}$ is $[\pmb{a}]_i$. The $N\times N$ identity, zero, and all-ones matrices are denoted by $\pmb{I}_N$, $\pmb{0}_{N\times N}$, and $\pmb{1}_{N\times N}$, respectively, and the $N$-dimensional zero and all-ones vectors by $\pmb{0}_N$ and $\pmb{1}_N$. The notation $\pmb{X}\succeq 0$ indicates that $\pmb{X}$ is positive semidefinite. The operators $\mathrm{diag}(\cdot)$ and $\mathrm{blkdiag}(\cdot)$ form diagonal and block-diagonal matrices from their arguments. For a matrix $\pmb{A}$, $\pmb{A}^T$, $\pmb{A}^H$, $\Re\{\pmb{A}\}$, and $\Im\{\pmb{A}\}$ denote the transpose, Hermitian transpose, real part, and imaginary part, respectively, and $\mathrm{Tr}(\pmb{A})$ denotes the trace. For a scalar $x$, $|x|$ is the absolute value, and $\|\cdot\|$ denotes the Euclidean norm for vectors or the Frobenius norm for matrices. The vectorization operator $\mathrm{vec}(\pmb{A})$ stacks the columns of $\pmb{A}$ into a single vector. The circularly symmetric complex Gaussian distribution with mean $\pmb{\mu}$ and covariance $\pmb{\Sigma}$ is denoted by $\mathcal{CN}(\pmb{\mu},\pmb{\Sigma})$, and the expectation operator by $\mathbb{E}[\cdot]$. For two real numbers $a$ and $b$, $(a,b)$ denotes the open interval between $a$ and $b$.
\section{System Model and Problem Formulation}
\begin{figure}[t]
    \centering
    \includegraphics[width=7cm]{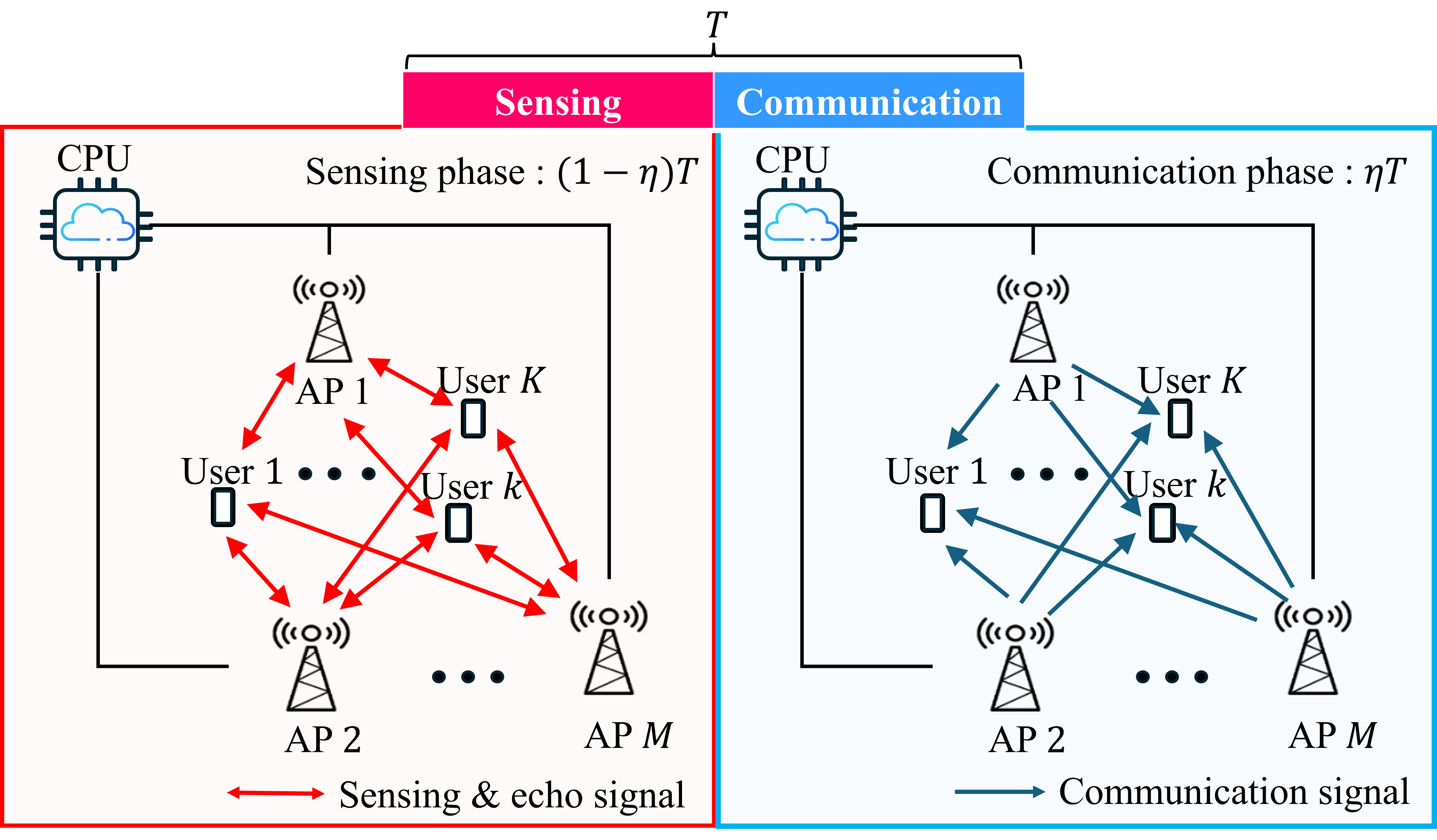}
    \caption{{T}ime-division cell-free near-field ISAC system.}
    \label{system model}
\end{figure}
As illustrated in Fig.~\ref{system model}, we consider a time-division cell-free near-field ISAC system, which consists of $M$ access points (APs) and $K$ single-antenna  users, indexed by $\mathcal{M} = (1, 2, \ldots, M)$ and $\mathcal{K} = (1, 2, \ldots, K)$, respectively. All APs are connected to a central processing unit (CPU) via ideal backhaul links with sufficiently large capacity~\cite{cellfree_backhaul}. Each AP operates in full-duplex mode with self-interference cancellation~\cite{cellfree_fD} and is equipped with $N_t = 2N+1$ uniform linear array (ULA) antennas, as shown in Fig.~\ref{fig:AP_channel}. 

The $M$ APs jointly perform user localization and simultaneously provide downlink communication services. The proposed time-division ISAC framework partitions each transmission block into sensing and communication phases, in which the time–frequency resources are jointly utilized for all users. Each transmission block of duration $T$ is partitioned into a sensing phase $T_{\text{s}}=(1-\eta)T$ and a communication phase $T_{\text{c}}=\eta T$, where $\eta \in (0,1)$ is the time allocation ratio \cite{time_allocation5, time_allocaion6}. The duration of $T$ is assumed to be short enough so that user positions—and thus the channels—remain nearly unchanged~\cite{time_allocation1, time_allocation2}. During the sensing phase, the system estimates the users’ locations and reconstructs the corresponding channels, which are then used for downlink communication in the subsequent communication phase. Moreover, the acquired location information can also be directly used to provide location-based services to the users. Compared to conventional approaches that perform sensing and channel estimation separately, 
the proposed integrated structure can significantly reduce pilot overhead, 
thereby improving overall performance~\cite{time_allocation1}. Moreover, by adjusting $\eta$, the system can flexibly balance the performance between sensing and communication according to their relative importance. 

The detailed channel, sensing, and communication models, along with the problem formulation, are provided in the following subsections.
\subsection{Channel Model}
\begin{figure}[t]
    \centering
    \includegraphics[width=7cm]{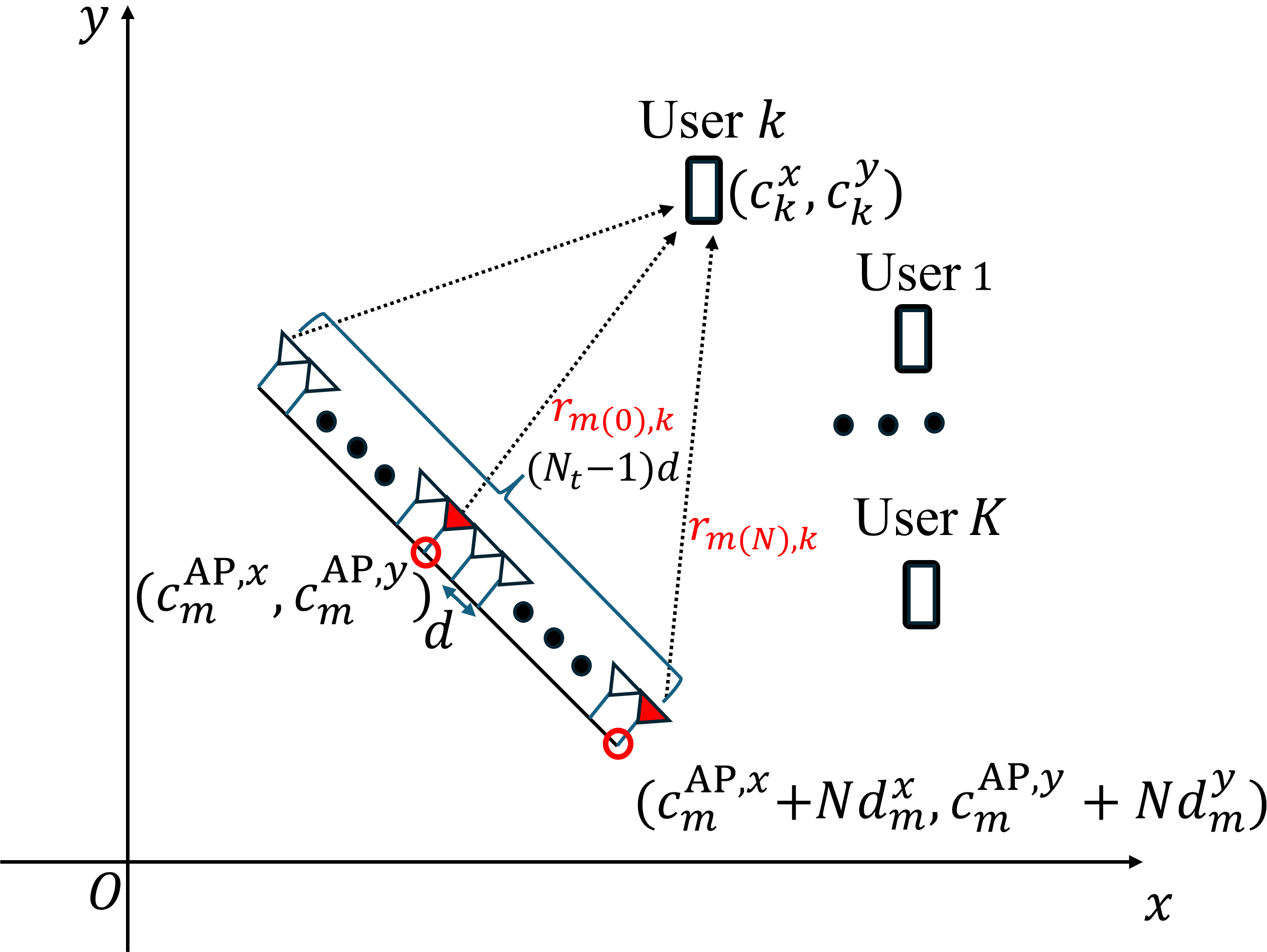}
    \caption{{Antenna configuration at AP $m$.}}
    \label{fig:AP_channel}
\end{figure}
We assume that all APs operate in full-duplex mode with circulators, enabling a shared antenna array to be used for both transmission and reception~\cite{full-duplex_technique}.
As illustrated in Fig.~\ref{fig:AP_channel}, both the ULAs and the users are assumed to be located on the $xy$-plane.\footnote{Similar to previous works that considered only 2D coordinates~\cite{nearfield_isac_1,nearfield_isac_2(velocity),nearfield_isac_3(multi-target)}, the proposed system can also be readily extended to the 3D case.}
 The distance between antenna elements of each AP is set to $d$, resulting in a ULA aperture of $D = (N_{\text{t}}-1) d$. The boundary distinguishing the near-field and far-field regions is determined by the Rayleigh distance, which is given by $2D^2/\lambda$ \cite{nearfield_review1}. 
 We consider a high-frequency band in the tens of gigahertz range and a large antenna array with hundreds of elements per AP, where the near-field region can span several hundred meters. Hence, all users are assumed to lie within the near-field region of all $M$ APs\cite{Nearfield_radar}.\footnote{If a specific user is in the far-field region of a particular AP, the distance between the AP and the user is sufficiently large. Due to the severe attenuation in mmWave frequencies, the user may not properly receive the signal. Consequently, we assume all users are in the near-field  region of all APs.}  In the high-frequency band under consideration,  non-line-of-sight (NLOS) components are significantly weaker than line-of-sight (LOS) path \cite{nearfield_isac_1,nearfield_isac_2(velocity),nearfield_isac_3(multi-target)}. Hence, only the LOS channel is considered in this work. Let the coordinates of the antenna center of AP $m$ be denoted by $(c_m^{\text{AP},x}, c_m^{\text{AP},y})$, 
the coordinates of user $k$ by $(c_k^x, c_k^y)$, 
and the coordinates of the $n$-th antenna element of AP $m$ by 
$(c_m^{\text{AP},x} + nd^{x}_m, \; c_m^{\text{AP},y} + n d_m^{y})$, 
for $n \in \{-N, \ldots, N\}$.
 Here, $d^{x}_m$ and $ d_m^{y}$ denote the $x$- and $y$-coordinate differences between adjacent antenna elements of AP $m$, satisfying $\sqrt{(d_m^{x})^2 +  (d_m^{y})^2} = d$.
Then, the distance from the $n$-th antenna of AP $m$ to user $k$ can be written as
\begin{align*}
r_{k,m(n)}\!=\! \sqrt{(c_k^x\! -\! (c_m^{\text{AP},x} + nd^{x}_m))^2\! +\! (c_k^y\! -\! (c_m^{\text{AP},y} + n d_m^{y}))^2}.
\end{align*}
Accordingly, the channel between the $n$-th antenna of AP $m$ and user $k$ can be expressed as \cite{tse2005fundamentals}
\begin{align*}
{h}_{k,m(n)} 
\!=\! \tilde{\beta}_{k,m(n)} e^{-j \frac{2\pi}{\lambda} r_{k,m(n)}} 
\!= \!\beta_{k,m(n)} e^{-j \frac{2\pi}{\lambda} (r_{k,m(n)}-r_{k,m(0)})}, 
\end{align*}
where $\beta_{k,m(n)} = \tilde{\beta}_{k,m(n)} e^{-j \frac{2\pi}{\lambda} r_{k,m(0)}}$ and $\tilde{\beta}_{k,m(n)} = \sqrt{\lambda/(4\pi r_{k,m(n)}^2)}$~\cite{nearfield_isac_1}. 
Since the variation of the channel gain is negligible compared to that of the phase shift \cite{nearfield_review1, nearfield_review2}, we assume $\tilde{\beta}_{k,m(n)} = \tilde{\beta}_{k,m(0)}$ for all $n$, and for notational simplicity we define $\beta_{k,m} = \beta_{k,m(0)}$.
Consequently, the near-field channel from AP $m$ to user $k$, 
$\pmb{h}_{k,m} \in \mathbb{C}^{N_t\times 1}$, can be expressed as
\begin{align}
\label{communication channel}
\hspace{-0.3cm}\pmb{h}_{k,m} \!=\!  
[{h}_{k,m(-N)}, \!\dots,\! {h}_{k,m(N)}]^T 
\!=\! \beta_{k,m} \pmb{a}_{k,m}\!(r_{k,m}^x, r_{k,m}^y\!),
\end{align}
where $r_{k,m}^x=c_m^{\text{AP},x}-c_k^x$, $ r_{k,m}^y=c_m^{\text{AP},y}-c_k^y$, and $\pmb{a}_{k,m}(r_{k,m}^x, r_{k,m}^y) \in \mathbb{C}^{N_t\times 1}$ 
denotes the array response vector from AP $m$ to user $k$ with its $n$-th element given by $
\left[\pmb{a}_{k,m}(r_{k,m}^x, r_{k,m}^y)\right]_n 
= e^{-j \frac{2\pi}{\lambda} \left(r_{k,m(n-N-1)} - r_{k,m(0)}\right)}$. 
As seen in~\eqref{communication channel}, accurate knowledge of user locations enables reconstruction of their communication channels. Using this observation, the proposed cell-free MIMO framework utilizes the user location information acquired during the sensing phase to directly infer the corresponding communication channels.

\subsection{Sensing Model}
The objective of sensing is to accurately determine the user's position. The mean square error (MSE) is widely used to evaluate the accuracy of the localization \cite{nearfield_isac_1}. However, obtaining a closed-form expression for such MSE is highly challenging. Therefore, we use the CRLB as a performance metric, which serves as a lower bound on the MSE\cite{nearfield_isac_1}. In the following, we first establish the echo signal model for the sensing phase and then derive the CRLB.

During the sensing phase, each AP continuously transmits sensing signals and receives their echoes to enable stable estimation. The number of sensing blocks, denoted by $\tau$, is determined by the sensing duration and given by $\tau = \left\lfloor (1 - \eta)T / \tau_{\text{s}} \right\rfloor$, where $\tau_{\text{s}}$ is the duration of a unit sensing block. Since $\tau$ must be an integer, the floor operation is applied; however, when $\tau_{\text{s}}$ is sufficiently small, the quantization effect is negligible. Hence, for analytical convenience, $\tau$ can be approximated as $\tau \approx (1 - \eta)T / \tau_{\text{s}}$. AP $m$ transmits a sensing signal $\pmb{s}_{m}[t]\in \mathbb{C}^{N_t \times 1}$ for sensing at slot $t$. Let the covariance matrix of $\pmb{s}_{m}[t]$ be denoted as $\pmb{\Psi}_{m} \in \mathbb{C}^{N_t \times N_t}$, i.e., $\pmb{\Psi}_{m} = \mathbb{E}[\pmb{s}_{m}[t] \pmb{s}_{m}[t]^H]$. Here, we assume that the sensing signals transmitted by AP $m$ and AP $m'$ are uncorrelated. That is, for $m \neq m'$, we have $\mathbb{E}[\pmb{s}_{m}[t] \pmb{s}_{m'}[t]^H] = \pmb{0}_{N_t\times N_t}$. 
{Therefore, the received echo at AP $l$ {at slot $t$ is} written as}
\begin{align*}
\label{receive sensing echo of l AP}
&\pmb{y}_l^{\text{AP}} [t]\! =\! \sum_{k=1}^{K}\! \sum_{m=1}^{M}\! \rho_{k,l,m}\pmb{h}_{k,l} (r_{k,l}^x,\! r_{k,l}^y\!) 
\pmb{h}_{k,m}^{H}\! (r_{k,m}^x\!,\! r_{k,m}^y\!) \pmb{s}_{m}[t]\! +\! \pmb{n}_l [t] \notag \\
&=\! \sum_{k=1}^{K}\! \sum_{m=1}^{M} \!\alpha_{k,l,m} \pmb{a}_{k,l} (r_{k,l}^x, \!r_{k,l}^y) 
\pmb{a}_{k,m}^{H}\!(r_{k,m}^x, \!r_{k,m}^y) \pmb{s}_{m} [t] + \pmb{n}_l[t],
\end{align*}
where $\rho_{k,l,m} \sim \mathcal{CN}(0,1)$ models the radar cross section (RCS)~\cite{cell-free4}, $\alpha_{k,l,m} = \rho_{k,l,m}\beta_{k,l}\beta_{k,m}$ represents the product of the user's RCS 
and the sensing-channel gains along the two-hop path 
AP $m$ $\rightarrow$ user $k$ $\rightarrow$ AP $l$, and $\pmb{n}_l[t] \sim \mathcal{CN}(0, \sigma^2 \pmb{I}_{N_t})$ is the additive Gaussian noise. After the sensing signal is reflected by the targets, since the CPU collects the received signals from all APs via high-capacity backhaul links, it can jointly perform sensing using the aggregated data \cite{cell-free1, cell-free2, cell-free3}. 
 Accordingly, the cascaded received signal vector is defined as
$
\pmb{y}^{\text{AP}}[t] = [\pmb{y}_1^{\text{AP}}[t]^T, \pmb{y}_2^{\text{AP}}[t]^T, \dots, \pmb{y}_M^{\text{AP}}[t]^T] \in \mathbb{C}^{MN_t \times 1},
$
which can be expressed as
\begin{align*}
\pmb{y}^{\text{AP}}[t] = 
\sum_{k}\pmb{H}_{k}(c^x_k, c^y_k, \pmb{\alpha}_k)\pmb{s}[t]
+
\pmb{n}[t],
\end{align*}
where $\pmb{\alpha}_k=(\alpha_{k,1,1}, \alpha_{k,1,2} \dots, \alpha_{k,M,M})$, $\pmb{s}[t]=[\pmb{s}_{1}[t]^T, \pmb{s}_{2}[t]^T,\dots,\pmb{s}_{M}[t]^T]^T \in \mathbb{C}^{MN_t \times 1}$ is the cascaded transmit sensing signal vector, $\pmb{n}[t] = [\pmb{n}_1[t]^T, \dots, \pmb{n}_M[t]^T]^T \in \mathbb{C}^{M N_t \times 1}$ 
denotes the stacked Gaussian noise vectors received at all APs, $\pmb{H}_{k}(c^x_k, c^y_k, \pmb{\alpha}_k)\in\mathbb{C}^{M N_t \times M N_t}$ is the composite sensing channel matrix for user $k$, structured as an $M \times M$ block matrix where each block size is 
$N_t \times N_t$. Let us use $\pmb{H}_{k}(c^x_k, c^y_k, \pmb{\alpha}_k)$ and $\pmb{H}_{k}$ interchangeably. 
The $(l,m)$-th block of $\pmb{H}_{k}$ is given by
$[\pmb{H}_{k}]_{\langle l,m\rangle}
= \alpha_{k,l,m}\,
\pmb{a}_{k,l}(r_{k,l}^x, r_{k,l}^y)\,
\pmb{a}_{k,m}^H(r_{k,m}^x, r_{k,m}^y),$
which represents the sensing channel from AP $m$, reflected by user $k$, 
and received at AP $l$. As echoed signals are received throughout the entire sensing phase of $\tau$ blocks, the aggregated received signal, $\pmb{Y}^{\text{AP}}\in \mathbb{C}^{MN_t \times \tau }$, can be expressed as follows:
\begin{align*}
\pmb{Y}^{\text{AP}} &= \begin{bmatrix} \pmb{y}^{\text{AP}}[1] & \dots & \pmb{y}^{\text{AP}}[\tau] \end{bmatrix} 
=\sum_{k}\pmb{H}_k(c^x_k, c^y_k, \pmb{\alpha}_k)\pmb{S}+\pmb{N},
\end{align*}
where $\pmb{S}=[ 
\pmb{s}[1],  \dots, \pmb{s} [\tau] 
] \in \mathbb{C}^{MN_t \times \tau}$ and $\pmb{N}=[\pmb{n}[1], \pmb{n}[2], \dots, \pmb{n}[\tau]] \in \mathbb{C}^{MN_t\times \tau}$. 
We assume that the number of blocks $\tau$ is sufficiently large, and the sample covariance matrix of $\pmb{s}[t]$ is approximated by the covariance matrix $\pmb{\Psi}\in \mathbb{C}^{MN_t \times MN_t}$. That is,  
$\pmb{\Psi} =  \mathrm{blkdiag}(\pmb{\Psi}_1, \dots, \pmb{\Psi}_M)=\mathbb{E}[\pmb{s}[t] \pmb{s}[t]^H]\approx \frac{1}{\tau} \pmb{S} \pmb{S}^H.$
To derive the CRLB, we vectorize the received signal matrix $\pmb{Y}^{\text{AP}}$. Then, the vectorized signal can be expressed as  
\begin{align*}
\pmb{u} = \text{vec}(\pmb{Y}^{\text{AP}}) \sim \mathcal{CN} \left( \text{vec}(\sum_k \pmb{H}_k(c^x_k, c^y_k, \pmb{\alpha}_k) \pmb{S}), \sigma^2 \pmb{I}_{MN_t\tau} \right).
\end{align*}
The unknown parameter vector to be estimated is given by  
$\pmb{\theta} = (\pmb{p}, \pmb{\alpha})$, where $\pmb{p}=(\pmb{c}^x, \pmb{c}^y)$ and $\pmb{\alpha}=( \pmb{\alpha}_{\text{re}}, \pmb{\alpha}_{\text{im}})$. Here, $\pmb{c}^x=(c_1^x,\dots, c_K^x)$, $\pmb{c}^y=(c_1^y, \dots, c_K^y)$, $\pmb{\alpha}_{\text{re}}=\Re\{(\pmb{\alpha}_{1}, \pmb{\alpha}_{2}, .., \pmb{\alpha}_{K} )\}$ and $\pmb{\alpha}_{\text{im}}=\Im\{(\pmb{\alpha}_{1}, \pmb{\alpha}_{2}, .., \pmb{\alpha}_{K} )\}$, respectively. The CRLB value of user position $\pmb{p}$ can be expressed as \cite{estimation_theory,CRLB}
\begin{align}
\text{CRLB}(\pmb{\Psi}, \eta) = \frac{\tau_{\text{s}}\sigma^2}{2{(1 - \eta)T}}\operatorname{Tr}(\bar{\pmb{J}}_{\pmb{p}\pmb{p}} - \bar{\pmb{J}}_{\pmb{p}\pmb{\alpha}}^T \bar{\pmb{J}}_{\pmb{\alpha}\pmb{\alpha}}^{-1} \bar{\pmb{J}}_{\pmb{p}\pmb{\alpha}})^{-1},
\end{align}
where $\bar{\pmb{J}}_{\pmb{p}\pmb{p}}$ is the normalized FIM for the user positions, 
$\bar{\pmb{J}}_{\pmb{\alpha}\pmb{\alpha}}$ for the gain parameters (including RCS), 
and $\bar{\pmb{J}}_{\pmb{p}\pmb{\alpha}}$ for their coupling. Let $\overline{\text{CRLB}}(\pmb{\Psi})=\sigma^2\operatorname{Tr}(\bar{\pmb{J}}_{\pmb{p}\pmb{p}} - \bar{\pmb{J}}_{\pmb{p}\pmb{\alpha}}^T \bar{\pmb{J}}_{\pmb{\alpha}\pmb{\alpha}}^{-1} \bar{\pmb{J}}_{\pmb{p}\pmb{\alpha}})^{-1}/2$. The detailed derivations of the CRLB, including the explicit formulations of $\bar{\pmb{J}}_{\pmb{p}\pmb{p}}$, $\bar{\pmb{J}}_{\pmb{p}\pmb{\alpha}}$, and $\bar{\pmb{J}}_{\pmb{\alpha}\pmb{\alpha}}$, are provided in Appendix~\ref{appendixA}.

It is important to note that the CRLB value is a increasing function of $ \eta $. In other words, allocating more time to the communication phase leads to degraded sensing performance. We design the covariance matrix $\pmb{\Psi}$ so as to minimize the CRLB, thereby improving the fundamental limit of localization accuracy. Based on the optimized sensing covariance, the actual user localization is then carried out using the multiple signal classification (MUSIC) algorithm~\cite{MUSIC}. 
{Details of its operation are provided in Appendix~\ref{AppendixB}.}  
\subsection{Communication Model}
In the communication phase, the channel is estimated based on the sensing information from the sensing phase. However, perfect sensing is difficult to achieve in practice. By considering channel uncertainty, we model the actual channel from AP $m$ to user $k$ as \cite{channel_uncertainty}
\begin{align}
\pmb{h}_{k,m} = \hat{\pmb{h}}_{k,m} + \Delta \pmb{h}_{k,m}, \quad \|\Delta \pmb{h}_{k,m}\| \leq \epsilon_{k,m},
\end{align}
where $\hat{\pmb{h}}_{k,m} \in \mathbb{C}^{N_t \times 1}$ denotes the estimated channel from AP $m$ to user $k$, $\Delta \pmb{h}_{k,m}\in \mathbb{C}^{N_t \times 1}$ represents the corresponding channel estimation error, and $\epsilon_{k,m}$ denotes the upper bound on the norm of the error vector. 

{Intuitively, larger localization errors lead to more severe  channel estimation errors. However, it is challenging to analytically quantify the channel estimation error induced by localization errors, because the induced channel error depends not only on the error magnitude but also on its direction as shown in \eqref{communication channel}. Therefore, we investigate this relationship numerically. Specifically, we consider an AP located at $(0,0)$m and fix the user–AP distance to $r=25$m. 
The user position error is modeled as a random vector with a fixed magnitude
$\Delta_r = \sqrt{(c_k^x-\hat{c}_k^x)^2 +(c_k^y-\hat{c}_k^y)^2}$ and a random direction uniformly distributed over $[0,2\pi)$, 
where $(\hat{c}_k^x, \hat{c}_k^y)$ denotes the estimated position of user $k$.}
The corresponding channel error averaged over the random direction of position error is then given as
$\Delta_h = \mathbb{E}\!\left[\|\pmb{h}_{k,m}(c_k^x,c_k^y)-\pmb{h}_{k,m}(\hat{c}_k^x,\hat{c}_k^y)\|\right],
$
where the expectation is computed via $10^5$ Monte Carlo realizations.

\begin{figure}[!t]
    \centering
    \subfloat[]{%
        \includegraphics[width=0.23\textwidth]{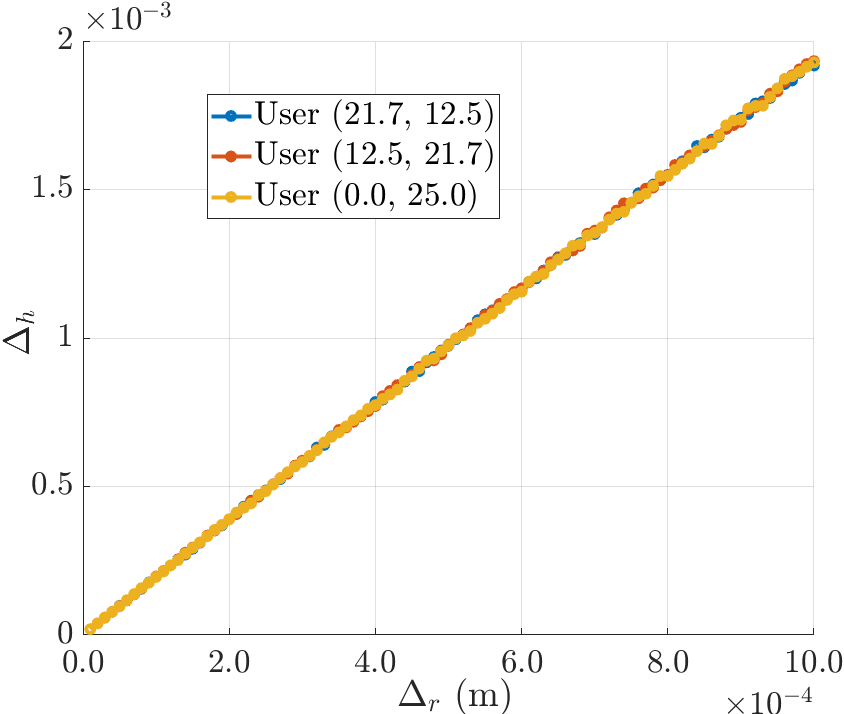}%
        \label{fig:channel_error_loc}%
    }\hfill
    \subfloat[]{%
        \includegraphics[width=0.23\textwidth]{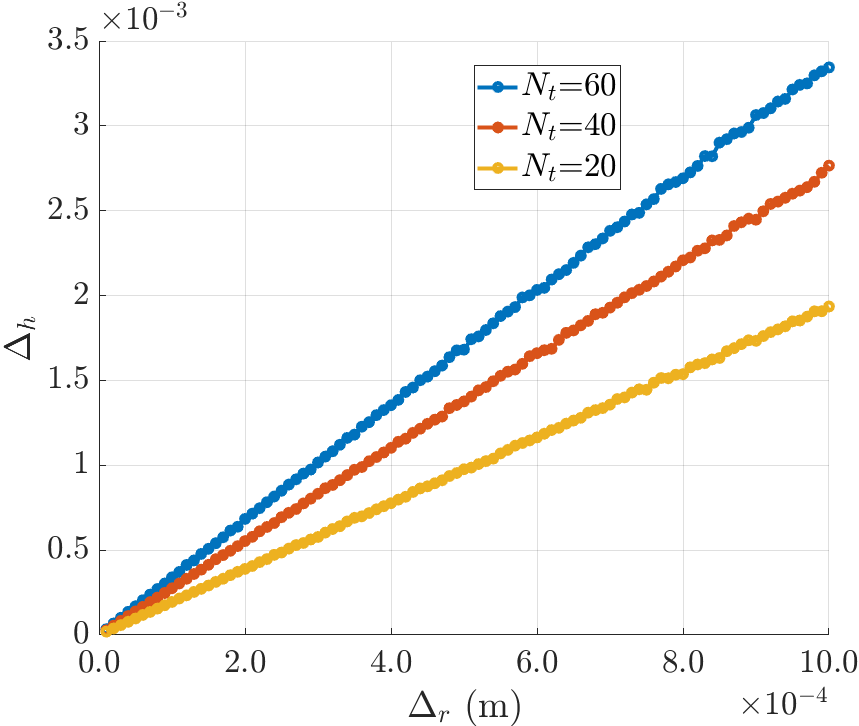}%
        \label{fig:channel_error_ant}%
    }
    \caption{Channel estimation error {versus} user localization error: (a) different user locations with fixed antenna number ($N_t=20$), (b) different antenna numbers with fixed user location ($c_k^x=12.5$, $c_k^y=21.7$).}
    \label{fig:channel_error_overall}
\end{figure}

As shown in Fig.~\ref{fig:channel_error_overall}(a), the channel error is plotted 
as a function of the localization error for three different user locations. 
Despite the varying positions, all curves exhibit nearly identical growth behavior, 
increasing monotonically with the localization error. 
This is because the AP-user distance is fixed to ensure channel-gain fairness, 
resulting in similar channel vector norms and thus similar error scaling.
In Fig.~\ref{fig:channel_error_overall}(b), the user location is fixed while the number of antennas varies over 20, 40, and 60. In this case, although the dependence on localization error follows a similar trend, the expected channel error $\Delta_h$ becomes 
larger for arrays with more antennas due to the increased channel vector norms 
(i.e., higher beamforming gain).
These results confirm that the channel error grows approximately linearly with the 
localization error and is proportional to the channel norm.

{Since the CRLB is closely related to the localization MSE, we define the following channel uncertainty bound:}
\begin{align}
\epsilon_{k,m}(\pmb{\Psi},\eta) = \alpha_{\text{e}} \sqrt{\operatorname{Tr}(\mathrm{CRLB}(\pmb{\Psi},\eta))}\, \|\pmb{h}_{k,m}\|, \label{eqn:estimation_error}
\end{align}
where $\alpha_{\text{e}}$ is a proportionality constant. It is worth noting that, instead of using the individual CRLB values of each user, 
the $\operatorname{Tr}(\mathrm{CRLB}(\pmb{\Psi},\eta))$, is used to provide an upper bound.

After obtaining the user's position and channel information through sensing, all APs transmit communication signals to the user. Let $\pmb{x}_m\in \mathbb{C}^{N_t \times 1}$ be the communication signal transmitted by AP $m$, then it can be expressed as $\pmb{x}_m = \sum_k \pmb{w}_{k,m} b_k$,
where $\pmb{w}_{k,m}\in \mathbb{C}^{N_t \times 1}$ denotes the transmit beamforming vector of AP $m$ for user $k$ and $b_k \sim \mathcal{CN}(0,1)$ denotes the information-bearing signal for user $k$. Let $\pmb{w} = \{\pmb{w}_{k,m}\}_{k \in \mathcal{K}, m \in \mathcal{M}}$. The received signal at user $k$ is given by
${y}_k = \sum_m \left(\pmb{h}_{k,m}^H \pmb{w}_{k,m} b_k+\sum_{k'\neq k}\pmb{h}_{k,m}^H \pmb{w}_{k',m} b_{k'}\right)+n_k,$
where $n_k \sim \mathcal{CN}(0,\sigma^2)$ is the additive Gaussian noise. $b_k$'s are assumed to be mutually uncorrelated, i.e., $\mathbb{E}[b_k b_{k'}^*] = 0$ for $k \ne k'$. The achievable rate of user $k$ is given as \cite{cell-free3}
\begin{align}
\label{rate}
&\!R_k(\pmb{w},\eta)\! =\! \eta T\! \log_2\! \left(1 \!+\! \frac{|\sum_m \pmb{h}_{k,m}^H \pmb{w}_{k,m}|^2}{\sum_{k' \ne k}|\sum_m  \pmb{h}_{k,m}^H \pmb{w}_{k',m}|^2 + \sigma^2} \right) = \notag \\ 
&\eta T \log_2 \!\left(\!1 + \frac{|\sum_m(\hat{\pmb{h}}_{k,m}+\Delta \pmb{h}_{k,m})^H \pmb{w}_{k,m}|^2}{\sum_{k' \ne k}|\sum_m (\hat{\pmb{h}}_{k,m}+\Delta \pmb{h}_{k,m})^H \pmb{w}_{k',m}|^2 \!+\! \sigma^2} \!\right).
\end{align}
Note that the channel uncertainty bound \eqref{eqn:estimation_error} is closely related to the sensing accuracy. In the subsequent sections, the functional dependencies of rates, CRLB and error bound on $\pmb{\Psi}$, $\pmb{w}$, $\eta$  may be omitted for brevity when the context is clear.
\subsection{Problem Formulation}
The proposed system estimates user locations in a sensing phase, and the resulting information is used both to provide sensing services and to construct the communication channels for the subsequent communication phase. Our objective is to maximize the worst-case communication rate across all users, while guaranteeing a required level of localization accuracy for sensing services. The corresponding optimization problem is formulated as follows:
\begin{subequations}
\label{origianl_problem}
\begin{align}
&\max_{\pmb{\Psi}, \pmb{w}, \eta} \min_k \min_{\|\Delta \pmb{h}_{k,m}\| \leq\epsilon_{k,m}(\pmb{\Psi},\eta), \forall m} R_k  \label{original objective}\\
\text{s.t.} 
\quad &  \text{CRLB}(\pmb{\Psi},\eta) \leq \text{CRLB}_{\text{th}}, \label{original CRLB constraint} \\
& \sum_k \|\pmb{w}_{k,m}\|^2 \leq P_{\text{max}}  \label{original com_power constraint}, \quad \forall m, \\
& \text{Tr}(\pmb{\Psi}_{m}) \leq P_{\text{max}}^{\text{s}}, \label{original sensing_power constraint} \quad \forall m, \\
&\pmb{\Psi}_{m} \succeq 0, \quad \forall m, \label{originia_sensing positive semidefinite}\\
& 0 <\eta < 1 \label{original_time division},
\end{align}
\end{subequations}
where the constraint in \eqref{original CRLB constraint} ensures that the CRLB values for user location estimation remain below a predefined threshold $\text{CRLB}_{\text{th}}$ for sensing services, \eqref{original com_power constraint} and \eqref{original sensing_power constraint} are communication and sensing power constraints with $P_{\text{max}}$ and $P_{\text{max}}^{\text{s}}$, respectively, 
 and \eqref{originia_sensing positive semidefinite} guarantees the positive semidefiniteness of the sensing covariance matrix. Finally, \eqref{original_time division} ensures a valid time allocation between the sensing and communication phases.
An interesting aspect of Problem \eqref{origianl_problem} is that the user rate looks like to increase linearly with $ \eta $. However, as $ \eta $ increases, the sensing accuracy deteriorates, which leads to an increase in the error bound  $\epsilon_{k,m}$. As a result, the user rate is not a simple increasing function of $ \eta $, but rather exhibits a trade-off between sensing and communication.
\section{{Main Algorithm}}
{Problem~\eqref{origianl_problem} is non-convex due to the intricate coupling between sensing accuracy and communication rate, rendering it generally intractable to solve directly. 
To address this challenge, we propose an algorithm that adopts an alternating
optimization (AO) framework~\cite{BCD} and leverages the S-procedure~\cite{generalized_S_procedure}
together with semidefinite programming (SDP) relaxation techniques~\cite{convexopt}
to solve the resulting problem.
This algorithm is referred to as \emph{Time-division ISAC algorithm with robust beamforming and time allocation (\textbf{TD-ISAC-Main})}. The detailed procedure is given in the following.}
\subsection {Sensing Matrix Optimization}
First, the optimal sensing covariance matrix remains unchanged regardless of the time allocation $\eta$, since the channels of all users are assumed to remain constant over the entire time slot duration $T$. Therefore, the sensing covariance matrix optimization problem can be formulated as follows:
\begin{align}
\label{sensing beamforming optimization}
\min_{\pmb{\Psi}}  \overline{\text{CRLB}}(\pmb{\Psi}) \quad \text{s.t.} \quad \eqref{original sensing_power constraint}, \eqref{originia_sensing positive semidefinite}.
\end{align}
We reformulate {Problem \eqref{sensing beamforming optimization}} by following the approach proposed in \cite{sensingbeamforming_reform}, by using slack variable $\pmb{U}$  as shown below:
\begin{subequations}
\label{sensing beamforming_schure complement}
\begin{align}
&\min_{\pmb{\Psi}, \pmb{U}} \text{Tr}(\pmb{U}^{-1})\\
\text{s.t.}
&\begin{bmatrix} 
\bar{\pmb{J}}_{\pmb{p}\pmb{p}}-\pmb{U}  & \bar{\pmb{J}}_{\pmb{p}\pmb{\alpha}} \\ 
\bar{\pmb{J}}_{\pmb{p}\pmb{\alpha}}^T & \bar{\pmb{J}}_{\pmb{\alpha}\pmb{\alpha}}
\end{bmatrix} \succeq 0, \\
&\eqref{original sensing_power constraint}, \eqref{originia_sensing positive semidefinite}.
\end{align}
\end{subequations}
Since Problem \eqref{sensing beamforming_schure complement} is convex, it can be solved via CVX~\cite{cvx}.
\subsection{Communication Beamforming and Time Allocation Optimization}
 This subsection focuses on optimizing the communication beamforming vector $\pmb{w}$ and the time allocation factor $\eta$, given a sensing covariance matrix $\pmb{\Psi}$. Optimization problem can be formulated as
\begin{align}
\label{imperfect problem}
&\max_{\pmb{w}, \eta} \min_k \min_{\|\Delta \pmb{h}_{k,m}\| \leq \epsilon_{k,m}, \forall m} R_k  \quad \text{s.t.} \quad \eqref{original CRLB constraint}, \eqref{original com_power constraint}, \eqref{original_time division}.
\end{align} Problem \eqref{imperfect problem} is difficult to solve directly due to the coupling between the beamforming vectors $\pmb{w}$ and the time-allocation ratio $\eta$. 
{In particular, varying $\eta$ impacts the channel uncertainty bound, which in turn affects the robust beamforming design.} {In the following, we first describe the optimization of one variable while fixing the other, namely beamforming or time allocation. We then present the overall alternating optimization algorithm with a discussion of its convergence and computational complexity.}

\subsubsection{{Beamforming Optimization}} We first fix the $\eta$, and introduce a slack variable $R$ to reformulate the worst-case rate maximization problem into a more tractable form.
\begin{subequations}
\label{imperfect slack use}
\begin{align}
&\max_{\pmb{w}, R} R  \\
\text{s.t.} \quad 
&\min_{\|\Delta \pmb{h}_{k,m}\| \leq \epsilon_{k,m}, \forall m} R_k  \geq R,\quad  \forall k,\label{imperfect constraint} \\
& \eqref{original com_power constraint}, \eqref{original_time division}.
\end{align}
\end{subequations}
Problem \eqref{imperfect slack use} is challenging to solve due to the worst-case rate constraint introduced by channel uncertainty in \eqref{imperfect constraint}. To eliminate the $\min$ operation in the constraint of \eqref{imperfect constraint}, we reformulate it as follows:
\begin{subequations}
\label{reform_worst}
\begin{align}
&\!\frac{1}{{\gamma}\!} \left|\sum_m\!(\hat{\pmb{h}}_{k,m}\!+\!\Delta\! \pmb{h}_{k,m})^H\! \pmb{w}_{k,m}\right|^2\!-\ \notag \\ &\sum_{k' \ne k}\left|\sum_m\! (\hat{\pmb{h}}_{k,m}\!+\!\Delta\! \pmb{h}_{k,m}\!)^H\! \pmb{w}_{k',m}\right|^2\  -\sigma^2\geq0, \quad \\
&\|\Delta \pmb{h}_{k,m}\| \leq \epsilon_{k,m},\forall k,m,
\end{align}
\end{subequations}
where ${\gamma}=2^{R/\eta T}-1$. 
Define
$\Delta\pmb{h}_k=[\Delta\pmb{h}_{k,1}^{T}, \dots, \Delta\pmb{h}_{k,M}^{T}]^T$, $\hat{\pmb{h}}_k=[\hat{\pmb{h}}_{k,1}^{T}, \dots, \hat{\pmb{h}}_{k,M}^{T}]^T$ and $\pmb{w}_k = [\pmb{w}_{k,1}^T, \pmb{w}_{k,2}^T, \dots, \pmb{w}^T_{k,M}]^T$, so that \eqref{reform_worst} can be rewritten as
\begin{subequations}
\begin{align}
\label{S_procedure_error}
\hspace{-0.3cm}\Delta\pmb{h}_k^{H}\pmb{M}_k\Delta\pmb{h}_k \!+\! 2\,&\Re\!\{\Delta\pmb{h}_k^{H}\pmb{M}_k\hat{\pmb{h}}_k\}\!+\!\hat{\pmb{h}}_k^{H}\pmb{M}_k\hat{\pmb{h}}_k \!-\!\sigma^{2}\! \;\geq\! 0 , \\ &\|\Delta \pmb{h}_{k,m}\| \leq \epsilon_{k,m}, \quad \forall k,m,
\end{align}
\end{subequations}
where $\pmb{M}_k = ({1}/{{\gamma}})\pmb{W}_k - \sum_{k' \ne k} \pmb{W}_{k'}$, $\pmb{W}_k=\pmb{w}_k\pmb{w}_k^H$.
However, the inequality in~\eqref{S_procedure_error} still involves the condition 
$\|\Delta \pmb{h}_{k,m}\| \leq \epsilon_{k,m}$, which makes the constraint 
difficult to handle directly due to its infinite number of possible channel perturbations.
{To address this difficulty, we leverage the following theorem from~\cite[Theorem~1]{generalized_S_procedure}.} {\begin{theorem}
 \label{theorem}
\textbf{(Generalized S-procedure)}~\cite[Theorem~1]{generalized_S_procedure}:
Let 
$f_0(\pmb{x}) = \pmb{x}^H \pmb{A}_0 \pmb{x} + 2 \Re\{\pmb{b}_0^H \pmb{x}\} + c_0$
and 
$f_i(\pmb{x}) = \pmb{x}^H \pmb{A}_i \pmb{x} + 2 \Re\{\pmb{b}_i^H \pmb{x}\} + c_i$
for $i \in \{1,\dots,L\}$, 
where $\pmb{A}_i \in \mathbb{C}^{N \times N}$ is a Hermitian matrix, 
$\pmb{b}_i \in \mathbb{C}^{N \times 1}$, and $c_i \in \mathbb{R}$ for all $i = 0,\dots,L$, 
and $\pmb{x} \in \mathbb{C}^{N \times 1}$.  
The implication
\[
f_i(\pmb{x}) \leq 0,\; \forall i = 1,\dots,L \quad \Rightarrow \quad f_0(\pmb{x}) \leq 0,
\]
holds if there exist scalars $\lambda_i \geq 0$ such that
\begin{equation*}
\sum_{i=1}^{L} \lambda_i 
\begin{bmatrix}
\pmb{A}_i & \pmb{b}_i \\
\pmb{b}_i^H & c_i
\end{bmatrix}
-
\begin{bmatrix}
\pmb{A}_0 & \pmb{b}_0 \\
\pmb{b}_0^H & c_0
\end{bmatrix}
\succeq 0.
\label{eq:generalized_s_procedure}
\end{equation*}
\end{theorem}}
By applying Theorem~\ref{theorem}, the robust constraint in~\eqref{S_procedure_error} can be equivalently reformulated as
\begin{equation}
\label{eq:S_procedure_LMI}
\begin{bmatrix}
\pmb{M}_k + \pmb{\Lambda}_k & \pmb{M}_k \hat{\pmb{h}}_k \\
\hat{\pmb{h}}_k^H \pmb{M}_k & \hat{\pmb{h}}_k^H \pmb{M}_k \hat{\pmb{h}}_k +  \sigma^2 - \pmb{\lambda}_k^T \pmb{\epsilon}_k
\end{bmatrix} \succeq {0}, \lambda_{k,m} \geq 0, \forall k,m,
\end{equation}
where $\pmb{\Lambda}_k=\text{blkdiag}(\lambda_{k,1}\pmb{I}_{N_t}, \dots, \lambda_{k,M}\pmb{I}_{N_t})$, $\pmb{\lambda}_k=[{\lambda}_{k,1}, \dots, {\lambda}_{k,M}]^T$ and $\pmb{\epsilon}_k=[\epsilon_{k,1}^2 , \dots, \epsilon_{k,M}^2 ]^T$. The constraint in~\eqref{eq:S_procedure_LMI} is non-convex due to two factors: 
the quadratic form in $\pmb{w}_{k}$ and the coupling between the beamforming vector $\pmb{w}_{k}$ and the rate threshold $R$. 
To handle the first factor, we reformulate the beamforming vector into a covariance matrix representation, i.e., $\pmb{W}_{k} = \pmb{w}_{k}\pmb{w}_{k}^H$, and apply an SDP relaxation by omitting the nonconvex rank-one constraint~\cite{convexopt}. 
To address the second factor, we exploit the fact that $R$ serves as a common rate threshold for all users; as $R$ increases, the required transmit power grows monotonically. 
Fortunately, when $R$ is fixed, the optimization problem becomes a linear matrix inequality (LMI) problem. 
This monotonicity and convexity enable a bisection approach, where the resulting convex feasibility problem is solved for each fixed $R$. 
This relaxation-and-search framework allows efficient computation of the robust beamforming matrix. 
Let $\pmb{W} = \{\pmb{W}_{k}\}_{k \in \mathcal{K}}$ and  $\pmb{\lambda}=\{\pmb{\lambda}_{k,m}\}_{k \in \mathcal{K}, m \in \mathcal{M}}$. 
The optimization problem is written as
\begin{subequations}
\label{imperfect_estimzation original problem_eta_del}
\begin{align}
\quad &\text{find} \quad \pmb{W}, \pmb{\lambda}    \label{error_objective_eta_del}\\
\text{s.t.} \quad &\pmb{W}_{k,m} \succeq 0, \quad \forall k,m, \label{positive semidefinite}\\
    & \sum_k \text{Tr}(\pmb{W}_{k,m}) \leq P_{\text{max}}, \quad \forall m, \label{SDP com_power constraint}\\
    & \eqref{eq:S_procedure_LMI}. 
\end{align}
\end{subequations}
Therefore, the beamforming matrix $\pmb{W}$ can be efficiently obtained for a given $\eta$ using the bisection method.
{This procedure, referred to as the bisection method for robust beamforming optimization (\textbf{Bi-RBO}), is summarized in \textbf{Algorithm~1}. It begins by taking $\eta$, $\overline{\text{CRLB}}(\pmb{\Psi})$, $P_{\text{max}}$, $R^{\text{u}}$, and $\epsilon_b$ as inputs, and then initializing the iteration index and bisection variables by setting $l=0$, $R^{(0)}=R^{\text{u}}$, and $R^{\text{d}}=0$.} At each iteration, Problem~\eqref{imperfect_estimzation original problem_eta_del} is solved. 
If the problem is infeasible, the upper bound is updated as $R^{\text{u}} = R^{(l)}$; otherwise, the lower bound is updated as $R^{\text{d}} = R^{(l)}$, and the rate threshold is set to $R^{(l)} = (R^{\text{u}} + R^{\text{d}})/2$, where $l$ denotes the iteration index. 
This iterative process continues until convergence, yielding the corresponding beamforming matrix $\pmb{W}$ for the given $\eta$.
\begin{algorithm}[t]
\caption{{Bisection method for robust beamforming optimization (\textbf{Bi-RBO})}}
\begin{algorithmic}[1]
\STATE {\textbf{Input:} $\eta$, $\overline{\text{CRLB}}(\pmb{\Psi})$, $P_{\text{max}}$, $R^{\text{u}}$, $\epsilon_b$.}
\STATE {Set $l=0$,  $R^{(0)}=R^{\text{u}}$ and $R^{\text{d}}=0$.}
\WHILE{$|R^{(l)}-R^{(l-1)})/R^{(l)}| > \epsilon_b$ \OR $l=0$}
\STATE $l \leftarrow  l + 1$
        \STATE Solve Problem \eqref{imperfect_estimzation original problem_eta_del}.
        \IF{Problem \eqref{imperfect_estimzation original problem_eta_del} is infeasible}
        \STATE Update $R^{\text{u}} = R^{(l)}$, and $R^{(l)}=(R^{\text{u}}+R^{\text{d}})/2$
        \ELSE
        \STATE Update $R^{\text{d}} = R^{(l)}$, and $R^{(l)}=(R^{\text{u}}+R^{\text{d}})/2$
        \ENDIF
\ENDWHILE
\STATE \textbf{Output:} $\pmb{W}$
\end{algorithmic}
\end{algorithm}
\subsubsection{{Time Allocation Optimization}}
After obtaining $\pmb{W}$, we optimize $\eta$ to further improve system performance. However, this optimization remains challenging even when $\pmb{w}$ is fixed, because $\epsilon_{k,m}$ depend on $\eta$, making the convexity analysis of Problem \eqref{imperfect problem} highly difficult. To address this issue, we propose optimizing $\eta$ to enlarge the feasible set of {Problem \eqref{imperfect_estimzation original problem_eta_del}}, thereby guaranteeing improved performance. We refer to this approach as the \textit{constraint-margin maximization algorithm}, which is novel in that it optimizes the time allocation ratio not for direct objective maximization but for feasibility expansion. This design enables the system to effectively balance sensing and communication under uncertainty. The detailed procedure is presented below.

In Problem \eqref{imperfect_estimzation original problem_eta_del}, the only $\eta$-dependent constraint is the LMI in \eqref{eq:S_procedure_LMI}, which is difficult to manipulate directly. By applying the Schur complement \cite{schur}, we reformulate it as
\begin{align}
\label{eq:A_eta}
A_k(\eta) \ge 0, \quad \forall k,
\end{align}
where $A_k(\eta)=
 \hat{\pmb{h}}_k^H \pmb{M}_k(\eta) \hat{\pmb{h}}_k
 + \sigma^2
 - \pmb{\lambda}^T_k \pmb{\epsilon}_k(\eta)
 - \hat{\pmb{h}}_k^H \pmb{M}_k(\eta) \bigl(\pmb{M}_k(\eta) + \pmb{\Lambda}_k\bigr)^{-1} \pmb{M}_k(\eta) \hat{\pmb{h}}_k.
$ If $A_k(\eta)$ can be increased by optimizing $\eta$, the constraint becomes more relaxed, thereby enlarging the feasible set for the subsequent optimization of $\pmb{W}$. Therefore, we update $\eta$ by solving
\begin{align}
\label{max_A_final}
\eta = \arg\max_{0<\eta<1} \min_k A_k(\eta).
\end{align}
Since Problem \eqref{max_A_final} is nonconvex due to the complicated form of $A_k(\eta)$, we adopt a one-dimensional search over $\eta \in (0,1)$ with step size $\epsilon_\eta$.

\subsubsection{{Overall Algorithm}}
Following the standard AO framework, we iteratively optimize the beamforming matrix $\pmb{W}$ and the time allocation $\eta$. To analyze convergence, let $(\pmb{W}^{(l)}, \eta^{(l)})$ be a feasible solution of the $l$-th iteration. Then, \eqref{eq:A_eta} holds for $(\pmb{W}^{(l)}, \eta^{(l)})$. Because increasing $A_k$ relaxes the LMI constraint, the previous solution $\pmb{W}^{(l)}$ remains feasible for any updated $\eta^{(l+1)}$. Moreover, \textbf{Bi-RBO} yields the globally optimal beamforming matrix $\pmb{W}$ for a given $\eta$. Consequently, the updated solution $\pmb{W}^{(l+1)}$ achieves a rate no worse than $\pmb{W}^{(l)}$, leading to the non-decreasing sequence.
Since the achievable rate is upper-bounded (e.g., by the transmit power), the sequence converges, ensuring both monotonicity and convergence of the AO procedure between $\eta$ and $\pmb{W}$.

After the objective value converges, the CRLB constraint in \eqref{original CRLB constraint} need to be satisfied. Let $\bar{\eta}$ denote the value that meets \eqref{original CRLB constraint} with equality. Then, $\bar{\eta}$ can be written as
\begin{align}
\label{CRLB}
    \bar{\eta}=1-\frac{\tau_{\text{s}}\sigma^2 \operatorname{Tr}((\bar{\pmb{J}}_{\pmb{p}\pmb{p}} - \bar{\pmb{J}}_{\pmb{p}\pmb{\alpha}}^T \bar{\pmb{J}}_{\pmb{\alpha}\pmb{\alpha}}^{-1} \bar{\pmb{J}}_{\pmb{p}\pmb{\alpha}})^{-1})}{2T\text{CRLB}_{\text{th}}}.
\end{align}
Let $\eta^{(L)}$ denote the value of $\eta$ returned at the last iteration $L$.
Since \eqref{original CRLB constraint} enforces an upper bound $\eta \le \bar{\eta}$, 
the final value of the time-allocation ratio $\eta^{\mathrm{opt}}$ is chosen as
\begin{align}
\label{optimal_eta}
\eta^{\mathrm{opt}} = \min\{\bar{\eta},\, \eta^{(L)}\},
\end{align}
which corresponds to projecting $\eta^{(L)}$ onto the feasible set. If the (unconstrained)
solution $\eta^{(L)}$ violates the CRLB constraint,
the boundary value $\bar{\eta}$ is selected. Furthermore, when $\eta^{\text{opt}}$ coincides with $\bar{\eta}$, the previously computed solution $\pmb{W}^{(l)}$ may not satisfy the optimality conditions. To ensure feasibility, the algorithm performs one additional beamforming optimization step in such cases.
Finally, since the SDP relaxation returns a beamforming matrix $\pmb{W}$ instead of a vector $\pmb{w}$, the beamforming vector is extracted via eigenvalue decomposition, i.e., $\pmb{w}_{k} =\rho_{k}\sqrt{\lambda_{\text{max}}(\pmb{W}_{k}^{(l)})}\pmb{v}_{\text{max}}(\pmb{W}_{k}^{(l)})$, where $\lambda_{\text{max}}(\cdot)$, $\pmb{v}_{\text{max}}(\cdot)$ are the largest eigenvalue of matrix and its corresponding eigenvector and $\rho_{k}$ is scaling factor for feasibility, thereby completing the algorithm. The overall procedure of \textbf{TD-ISAC-Main} is summarized in \textbf{Algorithm 2}. \\
\begin{algorithm}[t]
\caption{{Time-division ISAC algorithm with robust beamforming and time allocation (\textbf{TD-ISAC-Main})}}
\begin{algorithmic}[1]
\STATE {\textbf{Input:} $\eta^{(0)}$, $\epsilon$, $\epsilon_{\eta}$, $\epsilon_b$, $R^{\text{u}}$,$P_{\text{max}}$, $P_{\text{max}}^{\text{s}}$ and $\text{CRLB}_{\text{th}}$.}
\STATE {Set $l=0$.}
\STATE Find $\overline{\text{CRLB}}(\pmb{\Psi})$ by solving Problem \eqref{sensing beamforming_schure complement}.
\WHILE{$|(\eta^{(l)}-\eta^{(l-1)})/\eta^{(l-1)}|>\epsilon$ \OR $l=0$}
\STATE $l \leftarrow l+1$
\STATE $\pmb{W}^{(l)}$$\leftarrow$\textbf{Bi-RBO}($\eta^{(l-1)}$, $\overline{\text{CRLB} }$$(\pmb{\Psi})$, $P_{\text{max}}$, $R^{\text{u}}$, $\epsilon_b$).
\STATE Update $\eta^{(l)}$ by solving Problem \eqref{max_A_final}.
\ENDWHILE
\STATE Find optimal $\eta^{\text{opt}}$ by using \eqref{optimal_eta}.
\IF{$\eta^{\text{opt}}=\bar{\eta}$}
\STATE $\pmb{W}^{(l)}$$\leftarrow$\textbf{Bi-RBO}($\eta^{(l-1)}$, $\overline{\text{CRLB} }$$(\pmb{\Psi})$, $P_{\text{max}}$, $R^{\text{u}}$, $\epsilon_b$). 
\ENDIF
\STATE {Find $\pmb{w}_{k}$ via eigenvalue decomposition of $\pmb{W}_{k}^{(l)}$}.
\STATE {\textbf{Output}: $\pmb{w}$, $\eta$}
\end{algorithmic}
\end{algorithm}
We analyze the computational complexity of \textbf{TD-ISAC-Main}.  
We employ the interior-point method in CVX, whose complexity scales as $\mathcal{O}(V^{3.5})$, where $V$ denotes the number of optimization variables. 
In \eqref{sensing beamforming_schure complement}, optimizing the sensing covariance matrix involves $M$ matrices of size $N_t \times N_t$; since $N_t \gg K$, the additional slack variable has relatively small dimension and its impact is negligible, so the complexity scales as $\mathcal{O}\bigl((M N_t^2)^{3.5}\bigr)$. 
For the communication beamforming design, there are $K$ matrices of dimension $M N_t \times M N_t$, leading to a complexity of $\mathcal{O}\bigl((K (M N_t)^2)^{3.5}\bigr)$. 
The time-allocation update in \eqref{max_A_final} is handled via a one-dimensional search with complexity $\mathcal{O}(1/\epsilon_{\eta})$; for each candidate $\eta$, computing ${A}_k(\eta)$ requires matrix inversion with complexity $\mathcal{O}\bigl(K(MN_t)^3\bigr)$, so the overall complexity of optimizing $\eta$ is $\mathcal{O}\bigl((1/\epsilon_{\eta}) K(MN_t)^3\bigr)$. 
Combining these terms, the overall complexity is $\mathcal{O}\bigl((M N_t^2)^{3.5} + I\bigl((K (M N_t)^2 )^{3.5} + (1/\epsilon_{\eta}) K(MN_t)^3\bigr)\bigr)$
where $I$ is the number of outer iterations. 
Although \eqref{max_A_final} involves a one-dimensional search, \textbf{Bi-RBO} is executed only once per iteration until convergence, which significantly reduces the burden compared with exhaustive search. 
Nevertheless, the resulting algorithm still entails relatively high computational complexity, motivating the two lower-complexity suboptimal schemes introduced in the next subsection.

\section{Low-Complexity Algorithms}
{In the main algorithm (\textbf{TD-ISAC-Main}), the sensing covariance matrix is computed in a single step and yields the global optimum without iteration. In contrast, the optimization of communication beamforming and the time allocation requires iterative updates. Consequently, this stage incurs higher computational cost, as it involves solving an SDP within each bisection search and AO iteration. In this section, we propose two suboptimal algorithms with lower complexity for communication beamforming and time allocation optimization.}
\subsection{Time-Division ISAC Algorithm with Error-Ignorant Beamforming and Time Allocation}
{The first suboptimal algorithm is called time-division ISAC algorithm with error-ignorant beamforming and time allocation (\textbf{TD-ISAC-EI}).
This algorithm assumes perfect CSI by setting $\Delta \pmb{h}_{k,m}=\pmb{0}_{N_t}$, thereby reducing the computational complexity by avoiding the alternating-optimization (AO) framework and the S-procedure that are required to account for channel errors. 
Consequently, \textbf{TD-ISAC-EI} attains performance close to \textbf{TD-ISAC-Main} when the sensing constraint is stringent (i.e., localization is sufficiently accurate).
However, its performance degrades under large sensing errors because $\eta$ and the beamforming vectors are optimized without accounting for channel uncertainty.
The detailed procedure of \textbf{TD-ISAC-EI} is described as follows.}
\begin{algorithm}[t]
\caption{Time-division ISAC algorithm with error-ignorant beamforming and time allocation (\textbf{TD-ISAC-EI})}
\begin{algorithmic}[1]
\STATE {\textbf{Input:} $\epsilon$, $P_{\text{max}}$, $P_{\text{max}}^{\text{s}}$, $\text{CRLB}_{\text{th}}$, $\tilde{R}^{\text{u}}$.}
\STATE Set $l=0$, $\tilde{R}^{(0)}=\tilde{R}^{\text{u}}$ and $\tilde{R}^{\text{d}}=0$.
\STATE Find $\overline{\text{CRLB}}(\pmb{\Psi})$ by solving Problem \eqref{sensing beamforming_schure complement}.
\STATE Find optimal ${\eta}$ by using \eqref{CRLB}.
\WHILE{$|\tilde{R}^{(l)}-\tilde{R}^{(l-1)})/\tilde{R}^{(l)}| > \epsilon$ \OR $l=0$}
\STATE $l \leftarrow  l + 1$.        
        \STATE Solve Problem \eqref{SOCP}.
        \IF{\eqref{SOCP} is infeasible}
        \STATE Update $\tilde{R}^{\text{u}} = \tilde{R}^{(l)}$, and $\tilde{R}^{(l)}=(\tilde{R}^{\text{u}}+\tilde{R}^{\text{d}})/2$
        \ELSE
        \STATE Update $\tilde{R}^{\text{d}} = \tilde{R}^{(l)}$, and $\tilde{R}^{(l)}=(\tilde{R}^{\text{u}}+\tilde{R}^{\text{d}})/2$
        \ENDIF
\ENDWHILE
\STATE {\textbf{Output}: $\pmb{w}$, $\eta$}
\end{algorithmic}
\end{algorithm}
{Therefore, Problem~\eqref{imperfect problem} is rewritten as}
\begin{align}
\label{perfect_estimzation original problem}
\max_{\pmb{w}, \eta} \min_k R_k,  \quad \text{s.t.} \quad
    \eqref{original CRLB constraint}, \eqref{original com_power constraint}, \eqref{original_time division}.
\end{align}
In this case, the achievable rate $R_k$ becomes a linear function of $\eta$. Therefore, it is natural to maximize $\eta$ in order to increase $R_k$. However, as $\eta$ increases, the CRLB monotonically increases; therefore, the optimal $\eta$ makes the CRLB constraint in \eqref{original CRLB constraint} tight (i.e., it holds with equality). Accordingly, the optimal $\eta$ is $\bar{\eta}$ given by \eqref{CRLB}.
We now proceed to the optimization of beamforming for the communication users. 
Although the user rate expression $R_k$ has been simplified, Problem \eqref{perfect_estimzation original problem} remains challenging due to its max--min structure. Similar with Problem \eqref{imperfect slack use}, we introduce a slack variable $\tilde{R}$ and reformulate the problem as
\begin{subequations}
\label{perfect_estimzation original problem_eta_del}
\begin{align}
&\max_{\pmb{w},\tilde{R}} \tilde{R} \label{no-error objective_eta_del}\\
\text{s.t.} \quad
&R_k \geq \tilde{R}, \quad \forall k, \label{ganna_R_k} \\
& \eqref{original com_power constraint}.
\end{align}
\end{subequations}
However, Problem \eqref{perfect_estimzation original problem_eta_del} remains non-convex due to the logarithmic function in \eqref{ganna_R_k}. To overcome this, we expand the logarithmic term and transform the rate constraint as follows:
\begin{align}
\label{MRT SINR}
|\hat{\pmb{h}}_k^H \pmb{w}_k|^2 &\geq \tilde{\gamma} \left( \sum_{k' \neq k} |\hat{\pmb{h}}_k^H \pmb{w}_{k'}|^2 + \sigma^2 \right),
\end{align}
where $\tilde{\gamma} = 2^{\tilde{R}/(\bar{\eta} T)} - 1$. {Let us define  $\pmb{t}_k = [|\hat{\pmb{h}}^H_k \pmb{w}_1|, \ldots, |\hat{\pmb{h}}_k^H \pmb{w}_{k-1}|, |\hat{\pmb{h}}_k^H \pmb{w}_{k+1}|, \ldots, |\hat{\pmb{h}}_k^H \pmb{w}_K|, \sigma]^T$. Using the fact that $|\hat{\pmb{h}}_k^H \pmb{w}_k|
= |\hat{\pmb{h}}_k^H \pmb{w}_k e^{j\phi}| 
= \Re\{\hat{\pmb{h}}_k^H \pmb{w}_k\}$, the expression in \eqref{MRT SINR} can be equivalently written as
\begin{align}
\label{SOCP perfect CSIT}
\Re\{\hat{\pmb{h}}_k^H \pmb{w}_k\} \geq \sqrt{\tilde{\gamma}}
\| \pmb{t}_k\|.
\end{align}}
Nonetheless, Problem \eqref{perfect_estimzation original problem} remains non-convex due to the coupling between $\tilde{R}$ and $\pmb{w}$. 
Fortunately, when $\tilde{R}$ is fixed, \eqref{SOCP perfect CSIT} becomes a standard second-order cone programming (SOCP) constraint. 
Similar to $R$ in \textbf{TD-ISAC-Main}, the variable $\tilde{R}$ serves as a common rate threshold for all users.
For a given $\tilde{R}$, the beamforming vector $\pmb{w}$ can be obtained by solving the following feasibility problem:
\begin{align}
\label{SOCP}
&\text{find} \quad {\pmb{w}} \ \quad \text{s.t.}\quad \eqref{SOCP perfect CSIT}, \eqref{original com_power constraint}.
\end{align}
If Problem \eqref{SOCP} is feasible, $\tilde{R}$ is increased; otherwise, it is decreased.
This procedure yields the desired beamforming vector through a bisection search, and the procedure of \textbf{TD-ISAC-EI} is summarized in \textbf{Algorithm 3}. 
We analyze the computational complexity of the \textbf{TD-ISAC-EI}. The system involves $K$ users, where each beamforming vector is given by $\pmb{w}_k \in \mathbb{C}^{M N_t \times 1}$. 
Accordingly, the computational complexity of \textbf{TD-ISAC-EI} scales as $\mathcal{O}\bigl((K M N_t)^{3.5}\bigr)$. 
This represents a substantial reduction compared to the \textbf{TD-ISAC-Main}, where the $N_t^2$ term dominates the complexity and is now replaced by $N_t$. 
In addition, the constraint-margin maximization and the iterative procedure for optimizing $\eta$ are no longer required, which further reduces the overall computational burden. {For the \textbf{TD-ISAC-EI}, since $\tilde{R}$ is computed by ignoring channel errors, an additional post-processing step is required to obtain the worst-case rate {considering channel uncertainty}. Given the beamforming vectors $\pmb{w}$ obtained from the \textbf{TD-ISAC-EI}, Problem~\eqref{imperfect_estimzation original problem_eta_del} reduces to an LMI feasibility problem {with} the optimization variable $\pmb{\lambda}$, which can be efficiently solved via a bisection search over the rate threshold $R$, thereby yielding the worst-case rate.}
\subsection{Time-Division ISAC Algorithm with Robust Power and Time Allocation under MRT}
{The second suboptimal algorithm is called time-division ISAC algorithm with robust power and time allocation under MRT (\textbf{TD-ISAC-MRT}).
In the \textbf{TD-ISAC-MRT}, we assume maximum-ratio transmission (MRT) for beamforming. As $N_t$ increases, the user channels become asymptotically orthogonal, which makes MRT near-optimal in the large-antenna regime~\cite{nearfield_review1}, therefore, serves as an effective suboptimal beamforming strategy. Under MRT, the remaining optimization variables reduce to the power allocation across the MRT beams and the time-allocation ratio, which are optimized to be robust against channel errors. Compared with \textbf{TD-ISAC-EI}, where the beamforming vectors are designed under the perfect-CSI assumption and can suffer a severe performance loss when the channel error is large, \textbf{TD-ISAC-MRT} explicitly accounts for channel uncertainty in the power and time allocation.
As a result, \textbf{TD-ISAC-MRT} provides improved robustness while retaining low computational complexity. The detailed procedure is given below.}

In the MRT, beamforming vectors are designed based on the estimated channels at each AP. Accordingly, they are defined as follows
$\pmb{w}_{k,m} = \frac{\hat{\pmb{h}}_{k,m}}{\|\hat{\pmb{h}}_{k,m}\|} \, \sqrt{P_{k,m}}, \quad \forall k,m,$
where $P_{k,m}$ denotes the power allocated to user $k$ at AP $m$. Let define {$\pmb{p}_k = [\sqrt{P_{k,1}}, \ldots, \sqrt{P_{k,M}}]^T$}. {Similar to \eqref{imperfect problem}, for given $\eta$, we reformulate the optimization problem by introducing a slack variable $\tilde{R}^{\text{MRT}}$ as follows:}
\begin{subequations}
\label{MRT_problem}
\begin{align}
&\max_{\pmb{p}, \tilde{R}^{\text{MRT}}} \tilde{R}^{\text{MRT}} \\
\text{s.t.} &\min_{\|\Delta \pmb{h}_{k,m}\| \leq \epsilon_{k,m}, \forall m} R_k(\pmb{w}_k^{\text{MRT}},\eta)  \geq \tilde{R}^{\text{MRT}},\quad  \forall k, \label{MRT_robustconstraint} \\
&\sum_k P_{k,m} \leq P_{\text{max}}, \quad \forall m \label{MRT_power_constraint},
\end{align}
\end{subequations}
where {$\pmb{{p}}=\{\pmb{p}_{k}\}_{k\in\mathcal{K}}$ and $\pmb{w}_k^{\text{MRT}}=[\frac{\hat{\pmb{h}}_{k,1}}{\|\hat{\pmb{h}}_{k,1}\|} \, \sqrt{P_{k,1}}, ..., \frac{\hat{\pmb{h}}_{k,M}}{\|\hat{\pmb{h}}_{k,M}\|} \, \sqrt{P_{k,M}}]^T$.}
By applying Theorem~\ref{theorem}, we can express \eqref{MRT_robustconstraint} as
\begin{align}
\begin{bmatrix}
\pmb{M}^{\text{MRT}}_k + \pmb{\Lambda}_k & \pmb{M}^{\text{MRT}}_k \hat{\pmb{h}}_k \\
\hat{\pmb{h}}_k^{\mathsf H} \pmb{M}^{\text{MRT}}_k & \hat{\pmb{h}}_k^{\mathsf H} \pmb{M}^{\text{MRT}}_k \hat{\pmb{h}}_k + \sigma^2 - \pmb{\lambda}_k^{\mathsf T} \pmb{\epsilon}_k
\end{bmatrix}  \succeq 0, \quad \notag \\
\lambda_{k,m} \geq 0,\ \forall k,m, \label{eq:S_procedure_LMI_MRT}
\end{align}
{where $\pmb{M}^{\text{MRT}}_k = -({1}/{\tilde{\gamma}^{\text{MRT}}})\pmb{W}_k^{\text{MRT}} + \sum_{k' \ne k} \pmb{W}_{k'}^{\text{MRT}}$,
$\pmb{W}_k^{\text{MRT}} = \pmb{w}_k^{\text{MRT}} (\pmb{w}_k^{\text{MRT}})^{H} = \pmb{A}_k \pmb{p}_k\pmb{p}_k^H \pmb{A}_k^H$ and $\pmb{A}_k = \text{blkdiag}(\tfrac{\hat{\pmb{h}}_{k,1}}{\|\hat{\pmb{h}}_{k,1}\|}, \ldots, \tfrac{\hat{\pmb{h}}_{k,M}}{\|\hat{\pmb{h}}_{k,M}\|})$.
Similar to~\eqref{eq:S_procedure_LMI}, the constraint \eqref{eq:S_procedure_LMI_MRT} is non-convex due to the quadratic form in $\pmb{p}_k$ and its coupling with the common rate threshold $\tilde{R}^{\text{MRT}}$. To tackle this, we reformulate the power-allocation vector into a covariance matrix representation, i.e., $\pmb{P}_k = \pmb{p}_k \pmb{p}_k^{\mathsf H}$, and apply an SDP relaxation by omitting the nonconvex rank-one constraint. In addition, we exploit the monotonic relationship between $\tilde{R}^{\text{MRT}}$ and the required transmit power, which enables a bisection-based search over $\tilde{R}^{\text{MRT}}$. When $\tilde{R}^{\text{MRT}}$ is fixed, the optimization problem reduces to an LMI problem, therefore, Problem~\eqref{MRT_problem} can be efficiently handled via the bisection method. The corresponding feasibility problem is written as
\begin{align}
&{\text{find}} \quad \pmb{P},  \pmb{\lambda} \quad  \text{s.t.} \quad \eqref{MRT_power_constraint}, \eqref{eq:S_procedure_LMI_MRT},\label{imperfect_estimzation MRT problem_eta_del} 
\end{align}}where $\pmb{{P}}=\{\pmb{P}_{k}\}_{k\in\mathcal{K}}$. By replacing line 5 (solving Problem \eqref{imperfect_estimzation original problem_eta_del}) in \textbf{Bi-RBO} with the MRT-based formulation (solving Problem \eqref{imperfect_estimzation MRT problem_eta_del}), the optimization variable and output change from the beamforming matrix $\pmb{W}$ to the power allocation matrix $\pmb{P}$, while all other steps of the algorithm remain identical. We refer to the MRT-based variant of \textbf{Bi-RBO} as \textbf{Bi-RPO-MRT}.

After obtaining the MRT-based beamforming vectors, we proceed to optimize $\eta$ to further enhance the overall system performance.
By substituting the MRT beamforming into \eqref{max_A_final}, the optimization of $\eta$ can be performed in the same manner as in \textbf{TD-ISAC-Main}.
Specifically, $\eta$ is optimized to maximize the feasible region of the constraint, following the constraint-margin maximization algorithm.
Therefore, the $\eta$ optimization problem can be formulated as
\begin{align}
\label{max_A_MRT}
\eta = \arg\max_{0 < \eta < 1} \; \min_k \, A_k^{\text{MRT}}(\eta),
\end{align}
where
$A_k^{\text{MRT}}(\eta)= \hat{\pmb{h}}_k^H \pmb{M}_k^{\text{MRT}}(\eta) \hat{\pmb{h}}_k
 + \sigma^2
 - \pmb{\lambda}_k^T \pmb{\epsilon}_k(\eta)
 - \hat{\pmb{h}}_k^H \pmb{M}_k^{\text{MRT}}(\eta) \bigl(\pmb{M}_k^{\text{MRT}}(\eta) + \pmb{\Lambda}_k\bigr)^{-1} \pmb{M}_k^{\text{MRT}}(\eta) \hat{\pmb{h}}_k $. {\eqref{max_A_MRT} can be solved by a one-dimensional search with step size $\epsilon_\eta$.}

Similar to \textbf{TD-ISAC-Main}, the power allocation and $\eta$ optimization are iteratively updated until convergence, and the optimal $\eta^{\text{opt}}$ is determined by \eqref{optimal_eta}. If $\eta^{\text{opt}} = \bar{\eta}$, the bisection procedure (\textbf{Bi-RPO-MRT}) is executed again, after which the algorithm terminates. The optimization procedure for \textbf{TD-ISAC-MRT} is summarized in \textbf{Algorithm~4}. {Finally, the power-allocation vector is obtained as
$\pmb{p}_{k} = \rho_{k}^{\text{MRT}} \sqrt{\lambda_{\max}\big(\pmb{P}_{k}^{(l)}\big)}\, \pmb{u}_{\max}\big(\pmb{P}_{k}^{(l)}\big)$,
where $\rho_{k}^{\text{MRT}}$ is a scaling factor.}
In the MRT beamforming case, the overall algorithmic flow remains the same as in \textbf{TD-ISAC-Main}, with differences arising only in the mathematical formulation. Therefore, to calculate the complexity of \textbf{TD-ISAC-MRT}, it is sufficient to recalculate the complexity for \eqref{imperfect_estimzation MRT problem_eta_del} and substitute it into the overall expression. With $K$ users and $M$ APs, the total number of optimization variables reduces to $KM^2$. Accordingly, the computational complexity of the \textbf{TD-ISAC-MRT} is $\mathcal{O}\bigl(I\bigl((KM^2)^{3.5} + (1/\epsilon_{\eta})\,K(MN_t)^3\bigr)\bigr)$. {Compared with the complexity of \textbf{TD-ISAC-EI} which is dominated by $(K M N_t)^{3.5}$, \textbf{TD-ISAC-MRT} has the dominant term given as ${1}/{\epsilon_{\eta}}K (M N_t)^{3}$. Therefore, except when the $\epsilon_{\eta}$ is very small or the number of users and APs is small, \textbf{TD-ISAC-MRT} has typically less complexity than \textbf{TD-ISAC-EI} and has the lowest complexity among the proposed schemes}. 

\begin{algorithm}[t]
\caption{{Time-division ISAC algorithm with robust power and time allocation under MRT (\textbf{TD-ISAC-MRT})}}
\begin{algorithmic}[1]
\STATE {\textbf{Input:} $\eta^{(0)}$, $\epsilon$, $\epsilon_{\eta}$, $\epsilon_b$, $R^{\text{u}}$,$P_{\text{max}}$, $P_{\text{max}}^{\text{s}}$ and $\text{CRLB}_{\text{th}}$.}
{\STATE Set $l=0$.}
\STATE Find $\overline{\text{CRLB}}(\pmb{\Psi})$ by solving \eqref{sensing beamforming_schure complement}.
\WHILE{$|(\eta^{(l)}-\eta^{(l-1)})/\eta^{(l-1)}|>\epsilon$ \OR $l=0$}
\STATE $l \leftarrow l+1$
\STATE $\pmb{P}^{(l)}$$\leftarrow$\textbf{Bi-RPO-MRT}($\eta^{(l-1)}$, $\overline{\text{CRLB} }$$(\pmb{\Psi})$, $P_{\text{max}}$, $R^{\text{u}}$, $\epsilon_b$).
\STATE Update $\eta^{(l)}$ by solving \eqref{max_A_final}.
\ENDWHILE
\STATE Find optimal $\eta^{\text{opt}}$ by using \eqref{optimal_eta}.
\IF{$\eta^{\text{opt}}$=$\bar{\eta}$}
\STATE $\pmb{P}^{(l)}$$\leftarrow$\textbf{Bi-RPO-MRT}($\eta^{(l-1)}$, $\overline{\text{CRLB} }$$(\pmb{\Psi})$, $P_{\text{max}}$, $R^{\text{u}}$, $\epsilon_b$). 
\ENDIF
\STATE {Find $\pmb{p}_{k}$ via eigenvalue decomposition of $\pmb{P}_{k}^{(l)}$}.
{\STATE \textbf{Output}: $\pmb{P}$, $\eta$}
\end{algorithmic}
\end{algorithm}
\section{Numerical Results}
{We} present various simulation results to demonstrate the effectiveness of the proposed framework and algorithms. We set the number of users to $K = 4$ and the number of APs to $M = 3$. Each AP is equipped with a ULA consisting of {$N_t = 21$ antennas}, each operating at a carrier frequency of $f_c = 28$~GHz. The array aperture is given as $D = 0.53$~m \cite{nearfield_isac_1, nearfield_ISAC} and the near-field region is approximately $2D^2/\lambda \approx 53$~m. The noise variance is given as $\sigma^2=-70$~dBm. In a square area of $70\,\mathrm{m} \times 70\,\mathrm{m}$, the APs are located at $(30,70)$m, $(10,20)$m, and $(60,30)$m, while the users are positioned at $(20,30)$m, $(25,40)$m, $(40,30)$m and $(35,60)$m in the $xy$-plane. We assume $T=100$ ms, $\tau_s = 1$ ms. Based on the numerical result in Fig.~\ref{fig:channel_error_overall}, the proportionality  constant $\alpha_{\text{e}}$ in \eqref{eqn:estimation_error} is set to  $373$. {In this section, for brevity, the \textbf{TD-ISAC-Main}, the \textbf{TD-ISAC-EI}, and the \textbf{TD-ISAC-MRT} algorithms are referred to as \textbf{Main}, \textbf{EI}, and \textbf{MRT}, respectively.}
\begin{figure}[t]
    \centering
    \includegraphics[width=6cm]{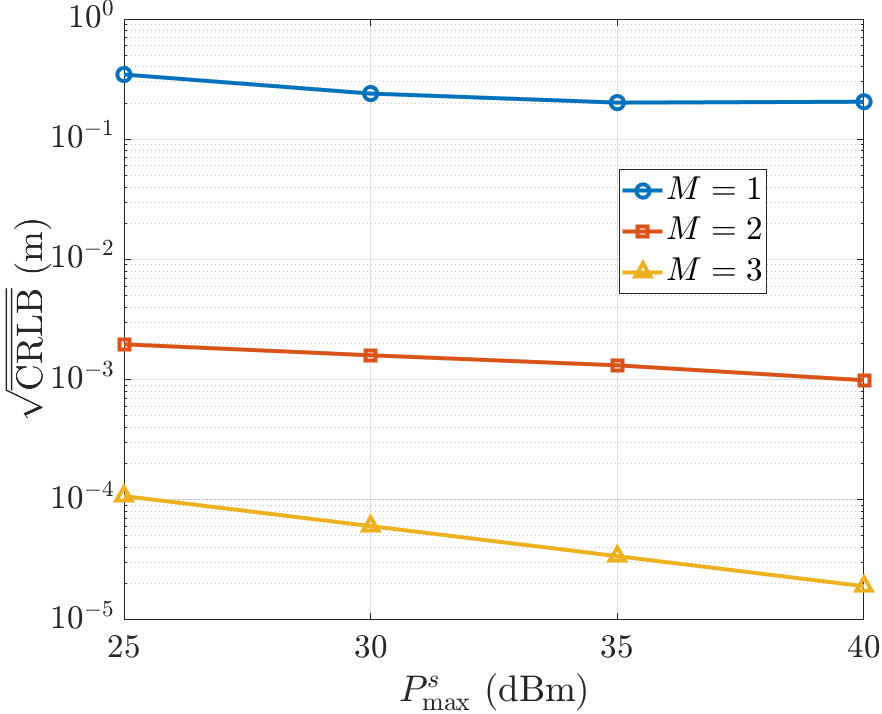}
    \caption{{$\sqrt{\overline{\text{CRLB}}}$ versus {$P_{\text{max}}^{\text{s}}$.}}}
    \label{fig:AP_power variation}
\end{figure}
\begin{figure*}[!t]
\centering
\subfloat[]{\includegraphics[width=0.23\textwidth]{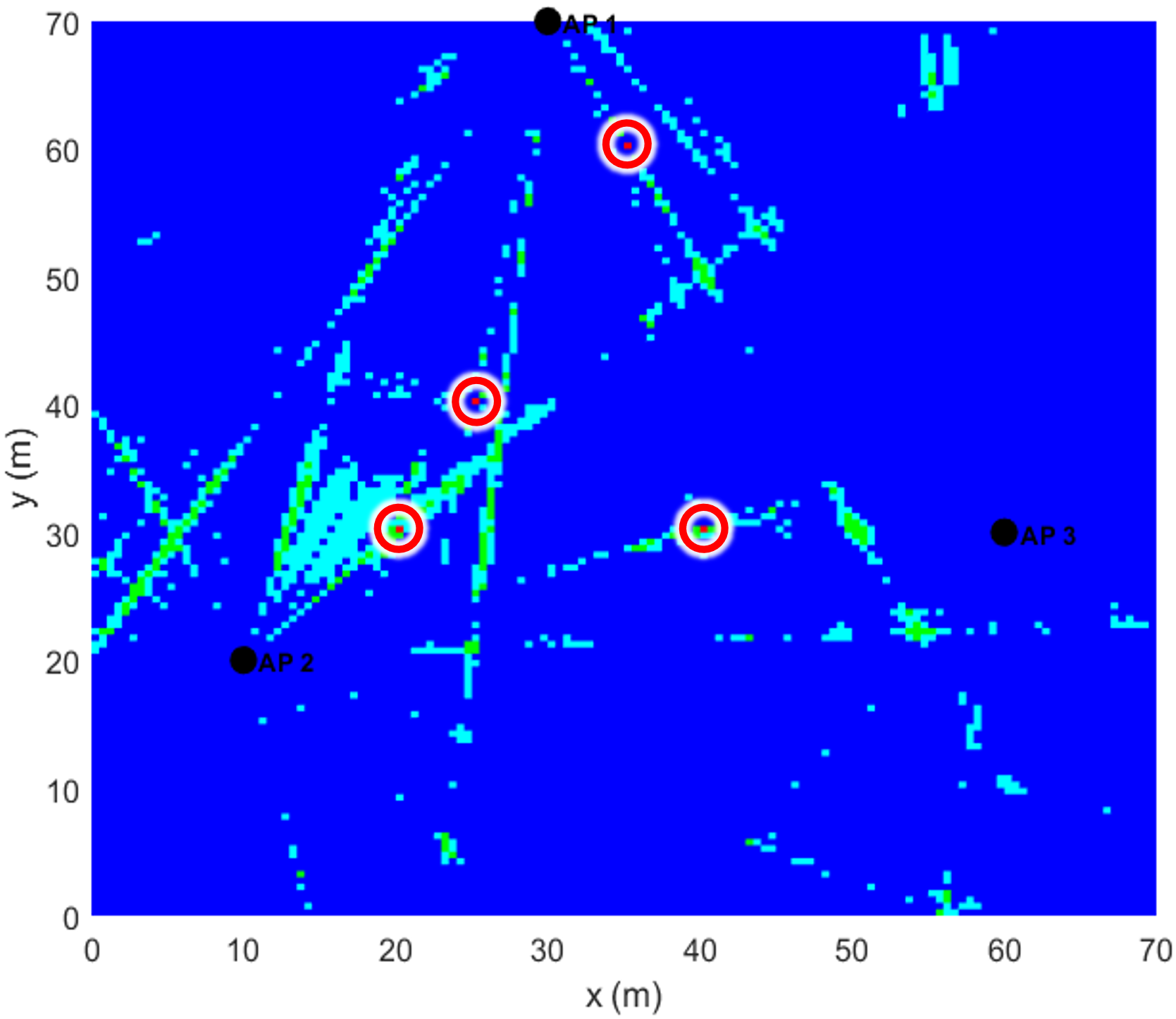}\label{fig:music_near}}\hfill
\subfloat[]{\includegraphics[width=0.23\textwidth]{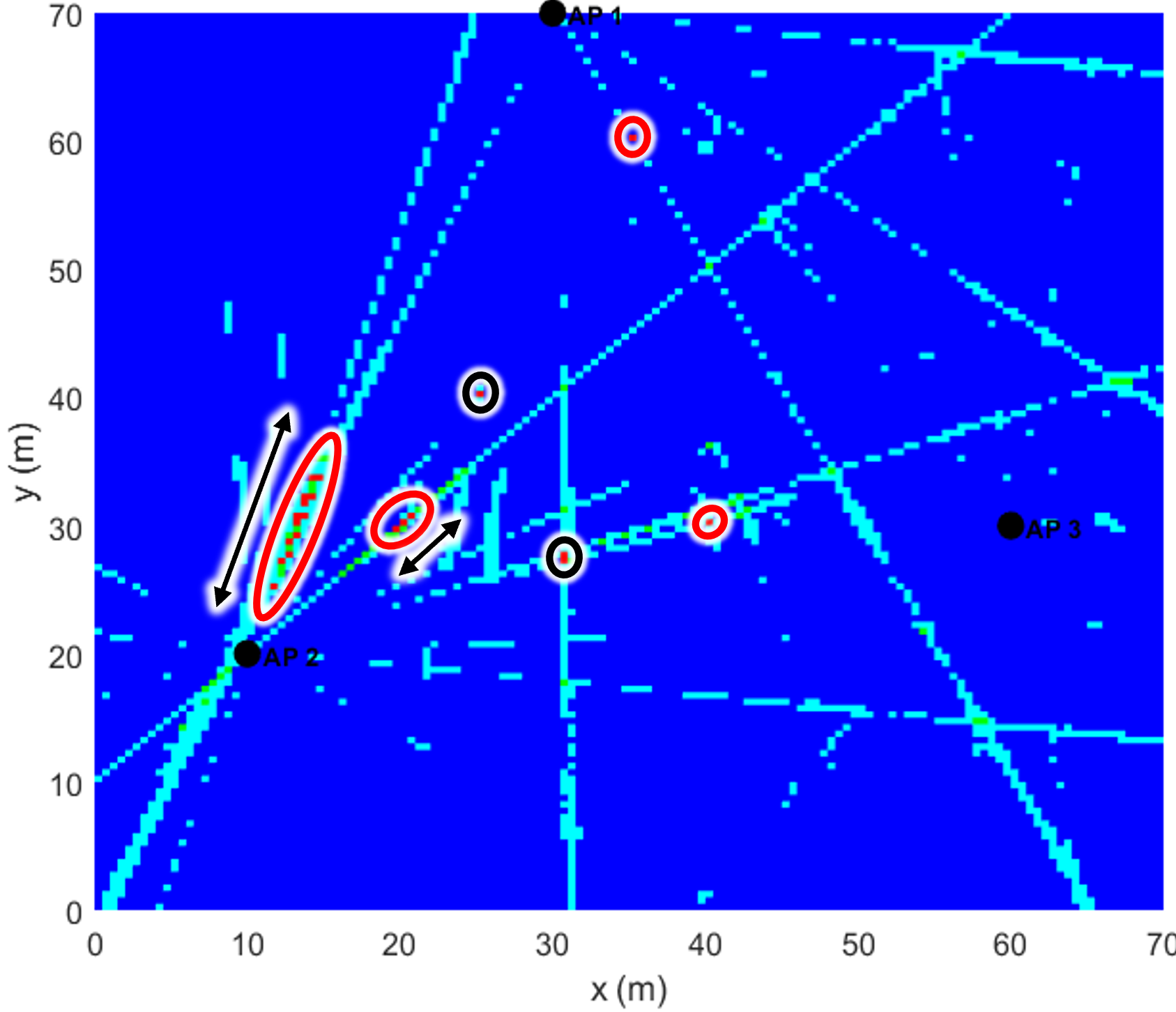}\label{fig:music_far}}\hfill
\subfloat[]{\includegraphics[width=0.23\textwidth]{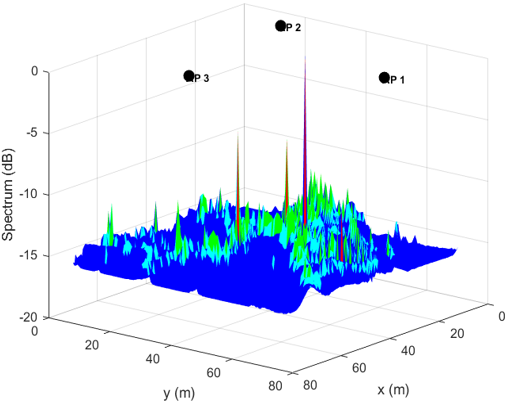}\label{fig:music_near_side}}\hfill
\subfloat[]{\includegraphics[width=0.23\textwidth]{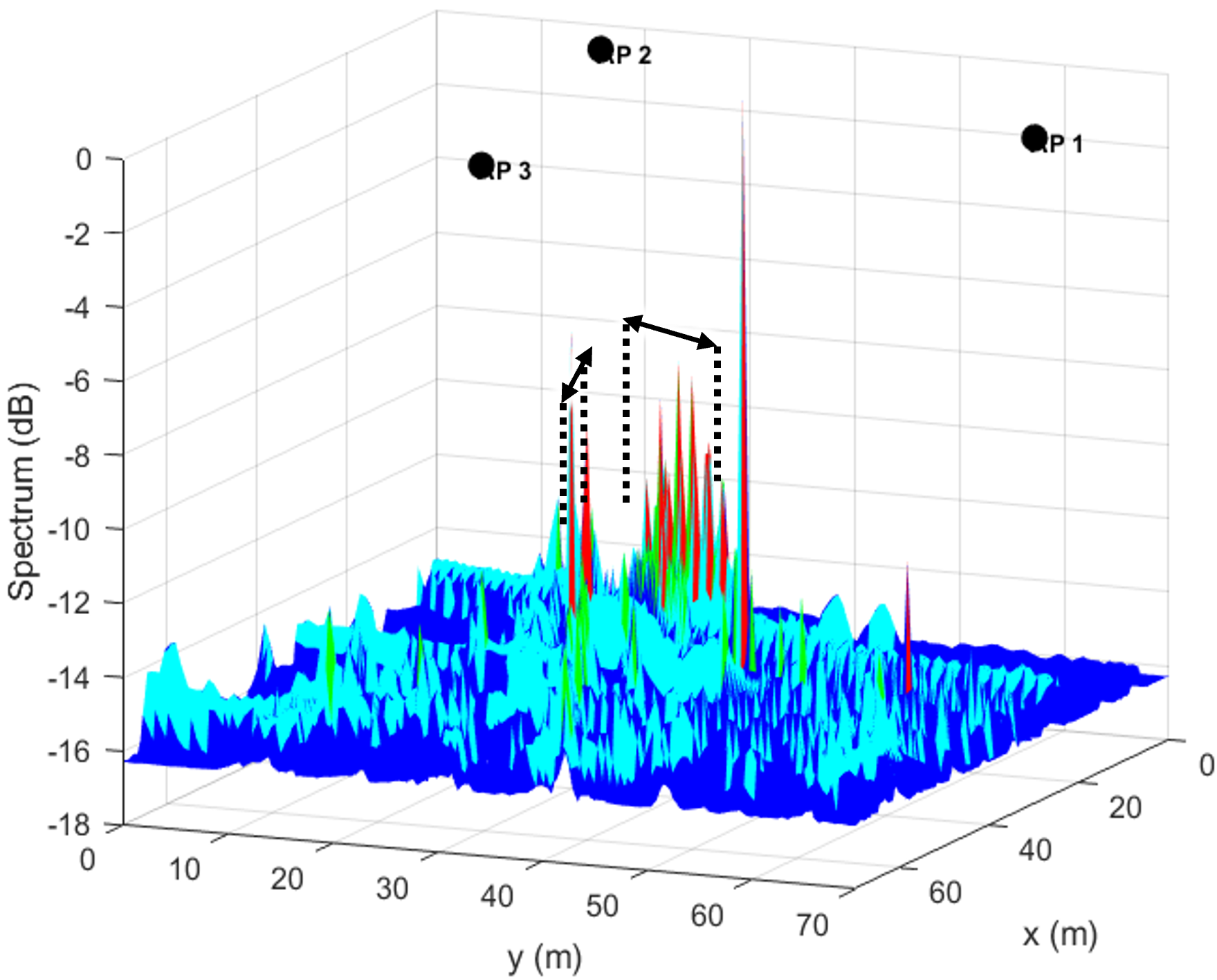}\label{fig:music_far_side}}
\caption{{Normalized MUSIC spectrum comparison between near-field and far-field scenarios with $P_{\text{max}}^{\text{s}}=40$ dBm.}}
\label{music}
\end{figure*}
\begin{figure}[t]
    \centering
    \includegraphics[width=5.9cm]{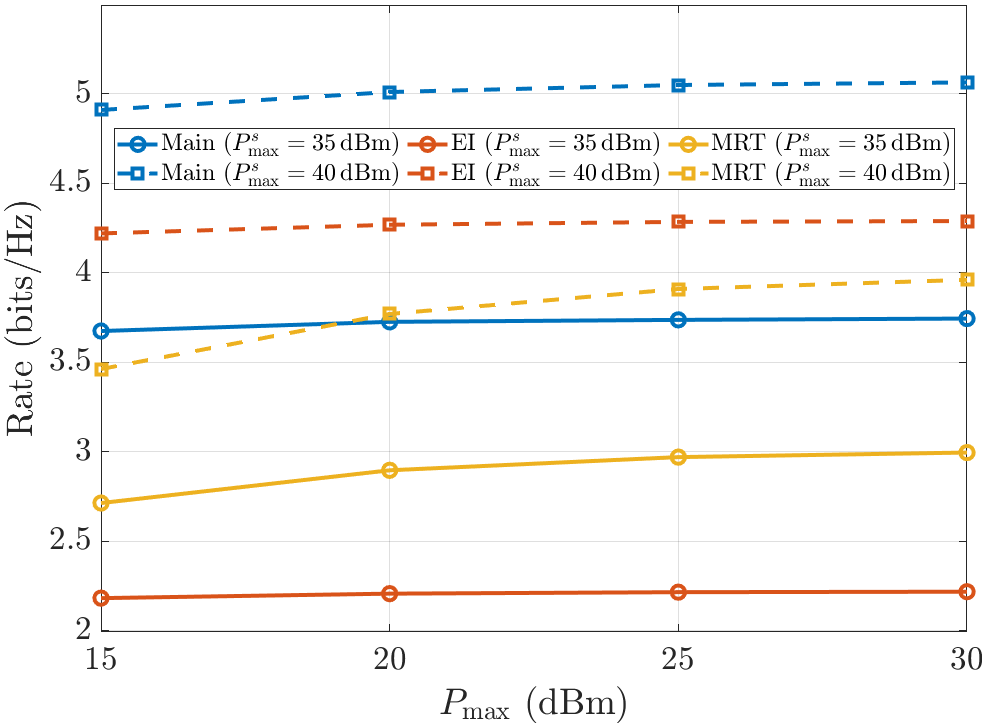}
    \caption{{Worst-user rate versus communication power for the proposed algorithms at two sensing power levels, with $\text{CRLB}_{\text{th}} = 1\times10^{-11}$.}}
    \label{fig:communicationcomparison1}
\end{figure}
\begin{figure}[t]
    \centering
    \includegraphics[width=5.9cm]{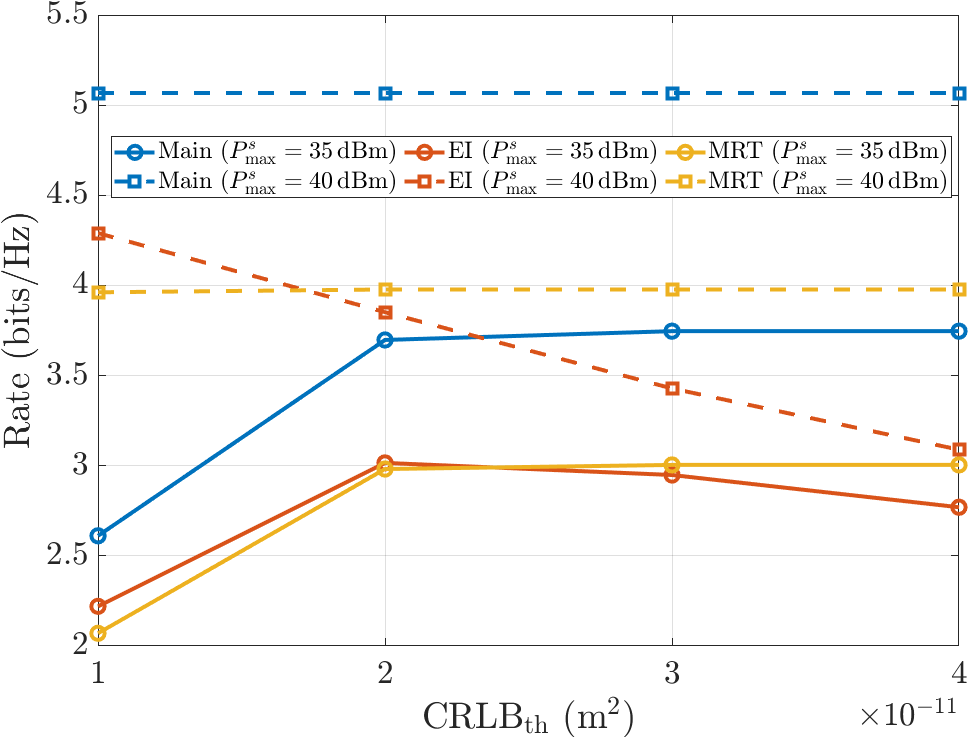}
    \caption{{Worst-user rate versus $\text{CRLB}_{\text{th}}$ for the proposed algorithms at two sensing power levels, when $P_{\text{max}} = 30$~dBm.}}
    \label{fig:communicationcomparison2}
\end{figure}

Fig.~\ref{fig:AP_power variation} illustrates $\sqrt{\overline{\text{CRLB}}}$ versus the sensing power for different numbers of APs. 
For fair comparison, when $M=1$ and $M=2$, each AP is equipped with 63 and 32 antennas, respectively, and their sensing power budgets are set to 3 and 1.5 times that of the $M=3$ case. 
As the sensing power increases, $\sqrt{\overline{\text{CRLB}}}$ decreases only moderately, whereas increasing the number of APs yields a much steeper and more consistent reduction, leading to significantly better localization performance than the single-AP baseline. 
This confirms that, in multi-user sensing scenarios, adding APs provides not only more antennas and higher total power, but also intrinsic multi-static gains such as enhanced channel diversity and spatial resolution.
Fig.~\ref{music} illustrates the sensing performance, where the 2D-MUSIC algorithm is applied using the optimized sensing covariance matrix $\pmb{\Psi}$ obtained by solving Problem~\eqref{sensing beamforming_schure complement}. 
The normalized MUSIC intensity is visualized with red for values above the minimum-intensity target, green for values within 5~dB below, cyan for 5–7~dB below, and blue for values more than 7~dB below this target. 
To emulate a far-field benchmark, the steering vector from AP~$l$ to user~$k$ is modeled as $\pmb{a}_{k,l}(r_{k,l}^x, r_{k,l}^y) = \pmb{a}_{k,l}(\theta_{k,l})$, where $\theta_{k,l}$ denotes the angle of departure (AoD) from AP~$l$ to user~$k$. As shown in Figs.~\ref{music}\subref{fig:music_near} and~\ref{music}\subref{fig:music_near_side}, the near-field MUSIC algorithm accurately localizes the users, with only the true user positions highlighted in red. 
In contrast, Figs.~\ref{music}\subref{fig:music_far} and~\ref{music}\subref{fig:music_far_side} show that, in the far-field case, not only the user positions but also several surrounding sidelobes reach the red level. This ambiguity arises because, in the far field, signals arriving from the same angle produce identical steering vectors and are therefore indistinguishable in the angular domain. 
For example, although the users at $(35,60)$ and $(40,30)$ are correctly detected, sidelobes near $(20,30)$ and $(25,40)$ produce higher peaks, resulting in localization errors. 
{These results show that the proposed near-field framework provides more accurate user localization and more reliable channel reconstruction than the far-field counterpart.}

{Fig.~\ref{fig:communicationcomparison1} illustrates {the worst-user rate of} the proposed algorithms under varying sensing and communication power budgets. As {either the communication power or the sensing power increases}, the achievable rates of all proposed schemes improve. In particular, increasing the sensing power leads to a substantially larger rate gain than increasing the communication power. In the presence of channel estimation errors, increasing the communication power offers only limited benefit because the residual channel error still constrains the performance, whereas increasing the sensing power directly reduces the estimation error magnitude and therefore yields much larger rate improvements.}
{Fig.~\ref{fig:communicationcomparison2} further illustrates this behavior under different CRLB thresholds and sensing power levels. 
At $P^{\text{s}}_{\text{max}} = 35$~dBm, increasing $\text{CRLB}_{\text{th}}$ from $1\times10^{-11}$ to $2\times10^{-11}$ improves the rate of all schemes. 
This is because at $P^{\text{s}}_{\text{max}} = 35$~dBm and $\text{CRLB}_{\text{th}} = 1\times10^{-11}$, the sensing constraint is stringent, which forces a small value of $\eta$ and thereby degrades the achievable rate.
For other parameter settings, however, varying $\text{CRLB}_{\text{th}}$ has little effect on the performance of \textbf{Main} and \textbf{MRT}, 
whereas \textbf{EI} continues to suffer a degradation in rate. 
This indicates that \textbf{Main} and \textbf{MRT} consistently adjust $\eta$ to balance communication and sensing performance, 
whereas \textbf{EI} determines $\eta$ solely to satisfy the sensing constraint with equality. 
Consequently, relaxing the sensing constraint increases the channel error in the \textbf{EI} scheme, which in turn leads to a reduction in the rate.}

\begin{figure}[!t]
    \centering
    \subfloat[]{%
        \includegraphics[width=0.23\textwidth]{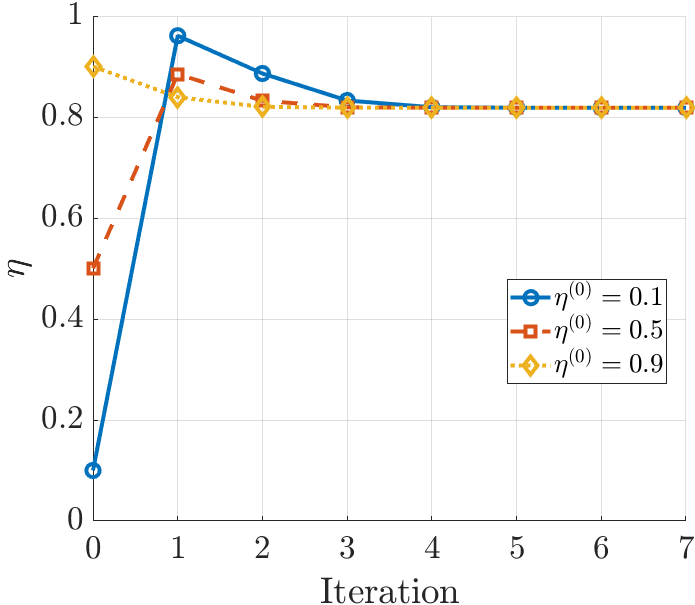}%
        \label{fig:eta_converge_eta}%
    }\hfill
    \subfloat[]{%
        \includegraphics[width=0.23\textwidth]{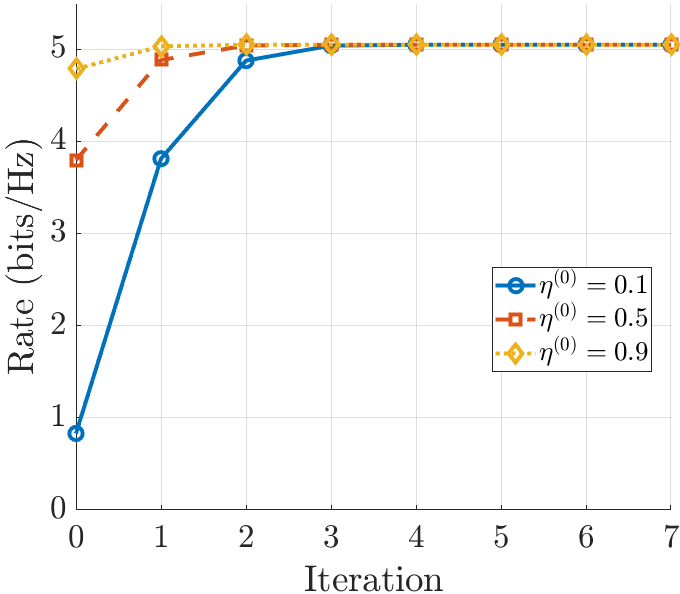}%
        \label{fig:eta_converge_rate}%
    }
    \caption{{Convergence of the \textbf{Main} algorithm: (a) $\eta$ for different initial values; (b) corresponding rate, when$P^{\text{s}}_{\text{max}} = 40$~dBm and $P_{\text{max}} = 30$~dBm.}}
    \label{fig:eta_convergence}
\end{figure}

\begin{figure}[t]
    \centering
    \includegraphics[width=5.3cm]{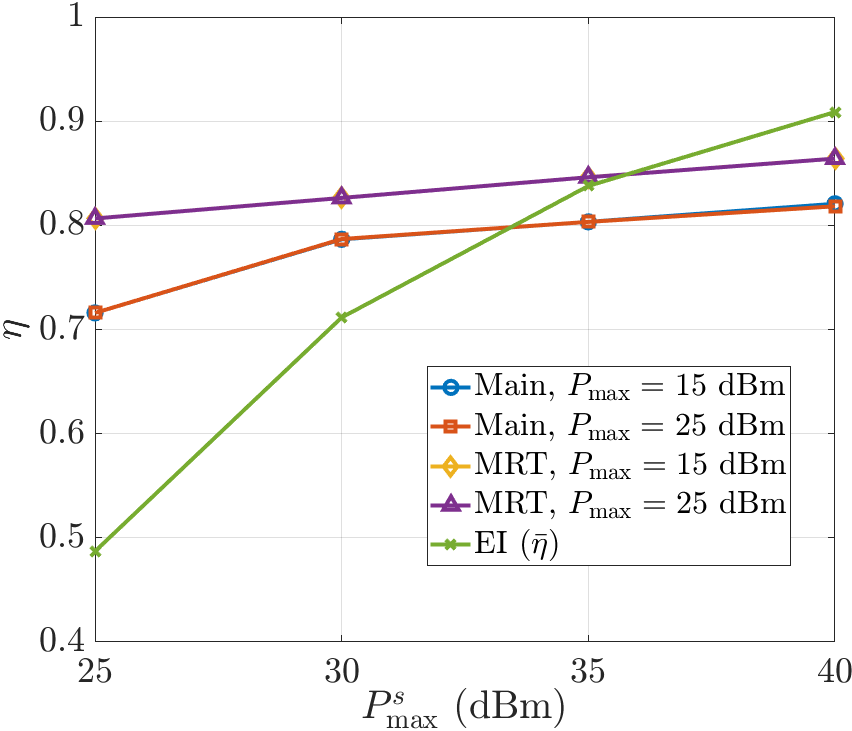}
    \caption{{Optimized time-allocation ratio $\eta$ versus sensing power $P^{\text{s}}_{\text{max}}$ {for the proposed algorithms at two communication power levels, when $\text{CRLB}_{\text{th}} = 2 \times 10^{-11}$.}}}
    \label{fig:eta_CRLB_var}
\end{figure}

Figs.~\ref{fig:eta_convergence}(a) and~\ref{fig:eta_convergence}(b) illustrate the convergence of the time-allocation ratio $\eta$ and the minimum user rate for the \textbf{Main} algorithm with initial values $\eta^{(0)} = 0.1$, $0.5$, and $0.9$. 
In all cases, the algorithm converges within four iterations to the same final $\eta$ and rate, regardless of the initialization. 
Although $\eta$ does not evolve monotonically, as seen in Fig.~\ref{fig:eta_convergence}(a), Fig.~\ref{fig:eta_convergence}(b) shows that the minimum rate increases monotonically. 
These results confirm that the \textbf{Main} algorithm is numerically stable and quickly converges to a good time-allocation ratio without requiring an exhaustive one-dimensional search.
Fig.~\ref{fig:eta_CRLB_var} shows how $\eta$ varies across different schemes as the sensing and communication powers change. 
As the sensing power increases, the sensing accuracy improves, so the CRLB requirement can be met with a shorter sensing duration $T_{\text{s}}$, and the time-allocation ratio $\eta$ consequently increases. 
In contrast, the communication power $P_{\text{max}}$ has little impact on $\eta$, whereas the choice of scheme does: \textbf{Main}, \textbf{MRT}, and \textbf{EI} yield different $\eta$ values due to their distinct $\eta$-optimization structures. 
In particular, \textbf{Main} and \textbf{MRT} arise from different formulations of the $\eta$-optimization problem, while \textbf{EI} selects $\eta$ by enforcing the CRLB constraint with equality.
 
\section{Conclusion}
{In this paper, we proposed a time-division near-field ISAC framework for cell-free systems, where sensing and downlink communication are jointly designed considering imperfect CSI due to localization errors. Numerical results demonstrate significant localization gains over mono-static and far-field baselines, while the suboptimal schemes achieve a favorable complexity–performance tradeoff: the error-ignorant design performs well under stringent sensing constraints, whereas the MRT-based design exhibits robustness to sensing requirements. These results confirm that near-field systems allow sensing to be exploited more effectively than conventional far-field architectures. In particular, in large-scale cell-free deployments, multi-static sensing effectively mitigates the channel-estimation bottleneck by significantly reducing channel uncertainty, thereby enabling a new location-aware channel-estimation paradigm for next-generation networks.}

{Future work will extend the framework to incorporate user mobility and dynamic resource allocation in heterogeneous edge environments. Another promising direction is to move beyond the purely sensing-based channel construction adopted in this work and develop hybrid channel-estimation schemes that jointly exploit pilot-based and sensing-based information, including algorithms that optimally combine the two estimates to further improve channel accuracy.}
\appendix
\subsection{CRLB Derivation}
\label{appendixA}
Following \cite{estimation_theory}, since $\pmb{u}$ is a Gaussian vector, the $(p,q)$-th element of the Fisher information matrix (FIM) $\pmb{J}$ with respect to $\pmb{\theta}$ is given by $[\pmb{J}]_{p,q} = \frac{2}{\sigma^2} \Re \left\{ \frac{\partial \pmb{u}^H}{\partial \theta_{p}} \frac{\partial \pmb{u}}{\partial \theta_{q}} \right\},$
where $\theta_{p}$ denotes the $p$-th element of the parameter vector $\pmb{\theta}$. Based on these observations, the FIM can be expressed in a block-matrix form.
\begin{align*}
\pmb{J} = \frac{2}{\sigma^2} \begin{bmatrix} \pmb{J}_{\pmb{p}\pmb{p}} & \pmb{J}_{\pmb{p}\pmb{\alpha}} \\ \pmb{J}_{\pmb{p}\pmb{\alpha}}^T & \pmb{J}_{\pmb{\alpha}\pmb{\alpha}} \end{bmatrix},
\end{align*}
where $\pmb{J}_{\pmb{p}\pmb{p}}$ denotes the FIM associated with the users' position, $\pmb{J}_{\pmb{\alpha}\pmb{\alpha}}$ corresponds to the FIM related to the gains, including radar cross section (RCS) and $\pmb{J}_{\pmb{p}\pmb{\alpha}}$ represents the FIM coupling between the user positions and gain parameters. For notational simplicity, all functional dependencies are omitted.

First, we analyze the $\pmb{J}_{\pmb{p}\pmb{p}}$. 
Since $\pmb{p}$ stacks the $x$- and $y$-coordinates of users, 
$\pmb{J}_{\pmb{p}\pmb{p}}$ has the following block-matrix form:
\begin{align*}
\pmb{J}_{\pmb{p}\pmb{p}} 
= \begin{bmatrix} 
\pmb{J}_{\pmb{c}^x\pmb{c}^x} & \pmb{J}_{\pmb{c}^x\pmb{c}^y} \\ 
\pmb{J}_{\pmb{c}^x\pmb{c}^y} & \pmb{J}_{\pmb{c}^y\pmb{c}^y} 
\end{bmatrix}.
\end{align*}
Since the channel of user $k' \neq k$ is independent of the position parameters of user $k$, 
the partial derivative of user $k'$s channel with respect to $(c_k^x, c_k^y)$ is zero.
Therefore, for ${q}, {r} \in \{x, y\}$, the $(k,k')$-th element of 
$\pmb{J}_{\pmb{c}^q\pmb{c}^r} \in \mathbb{R}^{K \times K}$ is given by
\begin{align*}
&[\pmb{J}_{\pmb{c}^q\pmb{c}^r}]_{k,k'} 
= \Re \left\{ \text{vec} \!\left( \frac{\partial \pmb{H}_k}{\partial c^q_k} \pmb{S} \right)^H 
\text{vec} \!\left( \frac{\partial \pmb{H}_{k'}}{\partial c_{k'}^r} \pmb{S} \right) \right\} \notag\\
&= \Re\{\text{Tr}(\pmb{S} \pmb{S}^H \tilde{\pmb{H}}_{c^q_k}^H \tilde{\pmb{H}}_{c_{k'}^r})\}
= \tau \, \Re\{\text{Tr}(\pmb{\Psi} \tilde{\pmb{H}}_{c^q_k}^H \tilde{\pmb{H}}_{c_{k'}^r})\},
\end{align*}
where $\tilde{\pmb{H}}_{c^q_k} = \partial \pmb{H}_k / \partial c^q_k$ denotes the partial derivative 
of the user $k$ channel matrix with respect to the $q$-coordinate of user $k$. The above holds because $\text{vec}(\pmb{A})^H \text{vec}(\pmb{A}) = \text{Tr}(\pmb{A}^H\pmb{A})$ 
and $\text{Tr}(\pmb{A}\pmb{B}\pmb{C}) = \text{Tr}(\pmb{C}\pmb{A}\pmb{B})$. In addition, $\pmb{H}_k \in \mathbb{C}^{MN_t \times MN_t}$ is partitioned into an $M \times M$ grid of blocks, 
each of size $N_t \times N_t$. 
The $(i,j)$-th block is given by 
$[\pmb{H}_k]_{\langle i,j\rangle} = \alpha_{k,i,j}\,\pmb{a}_{k,i}\pmb{a}_{k,j}^{\mathsf H}$. 
The partial derivative of this block with respect to $c_k^q$ is written as
\begin{align*}
&\left[\! \frac{\partial \pmb{H}_k}{\partial c^q_k}\! \right]_{<i,j>} \!
\!=\alpha_{i,j} \!\left(\! \frac{\partial \pmb{a}_{k,i}}{\partial c^q_k} \pmb{a}^H_{k,j}\! +\! \pmb{a}_{k,i} \frac{\partial \pmb{a}_{k,j}^H}{\partial c^q_k}\! \right).
\end{align*}
The derivative of the steering vector is
\begin{align*}
\frac{\partial \pmb{a}_{k,i}}{\partial c^q_k} \!\notag&=  \!
\left[ -j \tfrac{2\pi}{\lambda} \tfrac{-(c_i^{\text{AP},q} -Nd^q_{i} - c^q_k)}{\sqrt{(c_i^{\text{AP},x} - Nd^x_{i} - c_k^x)^2 + (c_i^{\text{AP},y} - Nd^y_{i} - c_k^y)^2}} a_{k,i(-N)},  \right. \\
&\!\dots,\! \left. -j \tfrac{2\pi}{\lambda} \tfrac{-(c_i^{\text{AP},q} + Nd^q_{i} - c^q_k)}{\sqrt{(c_i^{\text{AP},x} + Nd^x_{i} - c_k^x)^2 + (c_i^{\text{AP},y} + Nd^y_{i} - c_k^y)^2}} a_{k,i(N)} \right]^T,
\end{align*}
where $a_{k,i(n)}$ denotes the $n+N+1$-th element of $\pmb{a}_{k,i}$, defined as $a_{k,i(n)} = e^{-j \frac{2\pi}{\lambda} (r_{k,i(n)} - r_{k,i(0)})}$ for $n\in \{-N, \dots,N\}$.  

Next, we analyze the Fisher information matrix $\pmb{J}_{\pmb{\alpha}\pmb{\alpha}}$ associated with the path gains. The matrix $\pmb{J}_{\pmb{\alpha}\pmb{\alpha}}$ is also a block matrix and can be expressed as
\begin{align*}
\pmb{J}_{\pmb{\alpha}\pmb{\alpha}} = 
\begin{bmatrix}
    &\pmb{J}_{\pmb{\alpha}_{\text{re}}\pmb{\alpha}_{\text{re}}} \quad  & \pmb{0}_{KM^2 \times KM^2 } \\
    &\pmb{0}_{KM^2 \times KM^2} &\quad \pmb{J}_{\pmb{\alpha}_{\text{im}}\pmb{\alpha}_{\text{im}}}
\end{bmatrix},
\end{align*}
and $\pmb{J}_{\pmb{\alpha}_{\mathrm{re}}\pmb{\alpha}_{\mathrm{re}}}$, 
$\pmb{J}_{\pmb{\alpha}_{\mathrm{im}}\pmb{\alpha}_{\mathrm{im}}}\in \mathbb{R}^{KM^2 \times KM^2}$ matrices are written as
\begin{align*}
\pmb{J}_{\pmb{\alpha}_{\text{re}}\pmb{\alpha}_{\text{re}}} = \pmb{J}_{\pmb{\alpha}_{\text{im}}\pmb{\alpha}_{\text{im}}} =
\begin{bmatrix}
\!J_{\pmb{\alpha}_{1}\pmb{\alpha}_{1}} & J_{\pmb{\alpha}_{1}\pmb{\alpha}_{2}} & \cdots & J_{\pmb{\alpha}_{1}\pmb{\alpha}_{K}} \\
\!J_{\pmb{\alpha}_{2}\pmb{\alpha}_{1}} & J_{\pmb{\alpha}_{2}\pmb{\alpha}_{2}} & \cdots & J_{\pmb{\alpha}_{2}\pmb{\alpha}_{K}} \\
\!\vdots & \vdots & \ddots & \vdots \\
\!J_{\pmb{\alpha}_{K}\pmb{\alpha}_{1}} & J_{\pmb{\alpha}_{K}\pmb{\alpha}_{2}} & \cdots & J_{\pmb{\alpha}_{K}\pmb{\alpha}_{K}}
\end{bmatrix},
\end{align*}
where block matrix $\pmb{J}_{\pmb{\alpha}_k \pmb{\alpha}_{k'}}\in \mathbb{R}^{M^2 \times M^2}$ can be written as 
\begin{align}
\label{J_alpha}
\setlength{\arraycolsep}{3pt}    
\renewcommand{\arraystretch}{1}
\begin{bmatrix}
\!J_{\alpha_{k,1,1}\alpha_{k',1,1}}  & J_{\alpha_{k,1,1}\alpha_{k',1,2}} & \cdots & J_{\alpha_{k,1,1}\alpha_{k',M,M}} \\
\!J_{\alpha_{k,1,2}\alpha_{k',1,1}} & J_{\alpha_{k,1,2}\alpha_{k',1,2}} & \cdots & J_{\alpha_{k,1,2}\alpha_{k',M,M}} \\
\!\vdots & \vdots & \ddots & \vdots \\
\!J_{\alpha_{k,M,M}\alpha_{k',1,1}} & J_{\alpha_{k,M,M}\alpha_{k',1,2}} & \cdots & J_{\alpha_{k,M,M}\alpha_{k',M,M}}
\end{bmatrix}.
\end{align}
Since $\alpha_{k,i,j}$ is independent across different APs and users due to the random RCS, each element is defined as
\begin{align*}
J_{\alpha_{k,i,j}\alpha_{k',l,m}} 
= \tau \, \Re\left\{\operatorname{Tr} \!\left( 
\pmb{\Psi} (\pmb{D}_{i,j} \odot \bar{\pmb{H}}_k)^H 
(\pmb{D}_{l,m} \odot \bar{\pmb{H}}_{k'}) 
\right)\right\},
\end{align*}
where $\odot$ is Hadamard product, $\pmb{D}_{i,j} = \pmb{E}_{i,j} \otimes \pmb{1}_{N\times N}$ is a block-masking matrix, 
$\pmb{E}_{i,j} \in \mathbb{R}^{M \times M}$ is a selection matrix with a single one at the $(i,j)$-th position and zeros elsewhere, 
and $\bar{\pmb{H}}_k$ is the normalized channel matrix. 
Here, $\bar{\pmb{H}}_k$ is an $M \times M$ block matrix, where the $(i,j)$-th submatrix is defined as $[\bar{\pmb{H}}_k]_{<i,j>} = \pmb{a}_{k,i}\pmb{a}_{k,j}^H$.
Since $\mathbb{E}[\pmb{s}_{m}\pmb{s}_{m'}^H] = \pmb{0}_{N_t \times N_t}$ for $m \ne m'$, $\pmb{\Psi}$ is $M \times M$ block-diagonal matrix.
We can then exploit the trace property for block-diagonal matrices:
if $\pmb{A}$ is a block-diagonal matrix and $\pmb{B}$ is a block matrix with the same block dimensions as $\pmb{A}$,
$\operatorname{Tr}(\pmb{A}\pmb{B}) = \sum_i \operatorname{Tr}([\pmb{A}]_{<i,i>} \pmb{B}_{<i,i>})$. Therefore, 
\begin{align*}
\operatorname{Tr}\left( \pmb{\Psi} (\pmb{D}_{i,j} \odot \bar{\pmb{H}}_k)^H (\pmb{D}_{l,m} \odot \bar{\pmb{H}}_{k'}) \right) \notag
= \\
\sum_{r} \operatorname{Tr}\left([\pmb{\Psi}]_{<r,r>} \left[ (\pmb{D}_{i,j} \odot \bar{\pmb{H}}_k)^H (\pmb{D}_{l,m} \odot \bar{\pmb{H}}_{k'}) \right]_{<r,r>}\right).
\end{align*}
The diagonal entries of the matrix 
\(
(\pmb{D}_{i,j} \odot \bar{\pmb{H}}_k)^H (\pmb{D}_{l,m} \odot \bar{\pmb{H}}_{k'})
\) 
are nonzero only when the masking indices satisfy \( i = l \) and \( j = m \), in which case the nonzero entry appears at the \( (j,j) \)-th position.
Consequently, $\pmb{J}_{\pmb{\alpha}_k \pmb{\alpha}_{k'}}$ becomes a diagonal matrix, since it has nonzero entries only when $i = l$ and $j = m$. Therefore, \eqref{J_alpha} can be rewritten as
\begin{align*}
\pmb{J}_{\pmb{\alpha}_k\pmb{\alpha}_{k'}} \! = \! \operatorname{diag}\left( J_{\alpha_{k,1,1}\alpha_{k',1,1}}\!,\! J_{\alpha_{k,1,2}\alpha_{k',1,2}}\!,\! \dots\!,\! J_{\alpha_{k,M,M}\alpha_{k',M,M}}\! \right)\!,
\end{align*}
where each diagonal element is
$J_{\alpha_{k,i,j}\alpha_{k',i,j}}\! = \!\Re\{\operatorname{Tr}(\pmb{\Psi}_j 
(  [\bar{\pmb{H}}_k]_{<i,j>}^H [\bar{\pmb{H}}_{k'}]_{<i,j>} )\}$.

Finally, $\pmb{J}_{\pmb{p}\pmb{\alpha}}$ can be written as
\begin{align*}
\pmb{J}_{\pmb{p}\pmb{\alpha}}
=
\Re\!\left\{
\begin{bmatrix}
    \pmb{J}_{\pmb{c}^x\pmb{\alpha}_{\mathrm{re}}} &
    \pmb{J}_{\pmb{c}^x\pmb{\alpha}_{\mathrm{im}}} \\
    \pmb{J}_{\pmb{c}^y\pmb{\alpha}_{\mathrm{re}}} &
    \pmb{J}_{\pmb{c}^y\pmb{\alpha}_{\mathrm{im}}}
\end{bmatrix}
\right\},
\end{align*}
The matrices $\pmb{J}_{\pmb{c}^q\pmb{\alpha}_{\mathrm{re}}},\,\pmb{J}_{\pmb{c}^q\pmb{\alpha}_{\mathrm{im}}}\in\mathbb{R}^{K\times KM^{2}}$ 
collect the cross-information terms between user $q\in \{x,y\} $ coordinate and the real/imaginary channel coefficients. They can be written as 
\begin{align*}
&\pmb{J}_{\pmb{c}^q\pmb{\alpha}_{\text{re}}}\! =\! j\pmb{J}_{\pmb{c}^q\pmb{\alpha}_{\text{im}}} \!=\! \notag 
\begin{bmatrix}
\! J_{c^q_1 \alpha_{1,1,1}}\! &\! J_{c^q_1\alpha_{1,1,2}}\! &\! \cdots \!&\! J_{c^q_1\alpha_{K,M,M}}\! \\
\! J_{c^q_2 \alpha_{1,1,1}} & J_{c^q_2\alpha_{1,1,2}} & \cdots & J_{c^q_2\alpha_{K,M,M}} \\
\!\vdots & \vdots & \ddots & \vdots \\
\! J_{c^q_K \alpha_{1,1,1}} & J_{c^q_K\alpha_{1,1,2}} & \cdots & J_{c^q_K\alpha_{K,M,M}}
\end{bmatrix} ,
\end{align*}
with
\begin{align*}
&J_{{c}^q_k \alpha_{k',i,j}} = \tau \text{Tr}(\pmb{\Psi}\tilde{\pmb{H}}_{c^q_k}^H (\pmb{D}_{i,j} \odot \bar{\pmb{H}}_{k'}) )\notag \\
& =\!\tau \text{Tr}\! \left( \!\pmb{\Psi}_{j} \alpha_{i,j}^* \!\left(\! \frac{\partial \pmb{a}_{k,i}}{\partial c^q_k} \pmb{a}^H_{k,j}\! +\! \pmb{a}_{k,i} \frac{\partial \pmb{a}_{k,j}^H}{\partial c^q_k}\! \right)^H \!\pmb{a}_{k',i} \pmb{a}_{k',j}^H \right),
\end{align*}
where this equation follows since the $(j,i)$-th submatrix of $\tilde{\pmb{H}}_{c^q_k}^H$ and the $(i,j)$-th submatrix of $(\pmb{D}_{i,j} \odot \bar{\pmb{H}}_{k'})$ contribute to the $(j,j)$-th diagonal block.

According to \cite{estimation_theory}, the CRLB expression can be written as
\begin{align*}
\text{CRLB} 
&= \operatorname{Tr} \left( \left( \pmb{J}_{\pmb{p}\pmb{p}} - \pmb{J}_{\pmb{p}\pmb{\alpha}}^T \pmb{J}_{\pmb{\alpha}\pmb{\alpha}}^{-1} \pmb{J}_{\pmb{p}\pmb{\alpha}} \right)^{-1} \right) \notag \\&= \frac{\sigma^2}{2\tau}\operatorname{Tr}(\bar{\pmb{J}}_{\pmb{p}\pmb{p}} - \bar{\pmb{J}}_{\pmb{p}\pmb{\alpha}}^T \bar{\pmb{J}}_{\pmb{\alpha}\pmb{\alpha}}^{-1} \bar{\pmb{J}}_{\pmb{p}\pmb{\alpha}})^{-1} \notag \\
&=\frac{\tau_{\text{s}}\sigma^2}{2{(1 - \eta)T}}\operatorname{Tr}(\bar{\pmb{J}}_{\pmb{p}\pmb{p}} - \bar{\pmb{J}}_{\pmb{p}\pmb{\alpha}}^T \bar{\pmb{J}}_{\pmb{\alpha}\pmb{\alpha}}^{-1} \bar{\pmb{J}}_{\pmb{p}\pmb{\alpha}})^{-1}.
\end{align*}
where $\bar{\pmb{J}}_{ij} = ({\sigma^2}/{2\tau}) {\pmb{J}}_{ij}, \text{for } i,j \in \{\pmb{p},\pmb{\alpha}\}$. 
\subsection{MUSIC Algorithm}
\label{AppendixB}
MUSIC is a well-known localization algorithm that exploits the orthogonality between the signal and noise subspaces~\cite{MUSIC}. The detailed procedure of the MUSIC algorithm is presented as follows.
First, the CPU computes the sample covariance matrix $\bar{\pmb{\Psi}}$ of the received signal vector $\pmb{y}^{\text{AP}}[t]$ as  
\begin{align*}
\bar{\pmb{\Psi}} 
= \frac{1}{\tau}\pmb{Y}^{\text{AP}}(\pmb{Y}^{\text{AP}})^H
= \pmb{U}_{\text{s}} \pmb{\Lambda}_s \pmb{U}_{\text{s}}^H
+ \pmb{U}_{\text{n}} \pmb{\Lambda}_{\text{n}} \pmb{U}_{\text{n}}^H,
\end{align*}
where $\pmb{U}_{\text{s}}$ and $\pmb{U}_{\text{n}}$ denote the signal and noise subspace matrices, 
respectively, and $\pmb{\Lambda}_s$ and $\pmb{\Lambda}_{\text{n}}$ represent the diagonal matrices 
containing the signal and noise eigenvalues. 
Given $K$ sensing users, the signal subspace of the sensing echo covariance matrix $\bar{\pmb{\Psi}}$ 
is spanned by the eigenvectors associated with its $K$ largest eigenvalues. 
Since $\bar{\pmb{\Psi}}$ is constructed from the aggregated sensing echoes across all APs,
its signal subspace coincides with that generated by the composite steering vectors of all users.
Consequently, each column $\pmb{v}_k$ of $\pmb{U}_{\mathrm{s}}$ represents the steering
vector of user $k$, which can be written as
\begin{align*}
\pmb{v}_k 
= [\pmb{a}_{k,1}^T(r_{k,1}^x, \!r_{k,1}^y),\; \cdots,\; \pmb{a}_{k,M}^T(r_{k,M}^x, \!r_{k,M}^y)]^T.
\end{align*}
Accordingly, we define the composite steering vector $\tilde{\pmb{v}}(c^x, c^y)\in \mathbb{C}^{M N_t \times 1}$ as
\begin{align*}
\tilde{\pmb{v}}(c^x, c^y) =
[\pmb{a}_1^T(c^x, c^y), \cdots, \pmb{a}_M^T(c^x, c^y)]^T,
\end{align*}
where the $(n+N+1)$-th element of $\pmb{a}_m(c^x, c^y)$ is given by
\begin{align*}
[\pmb{a}_m(c^x, c^y)]_{n+N+1}
= e^{-j \frac{2\pi}{\lambda} \left(r_{m(n)}(c^x, c^y) - r_{m(0)}(c^x, c^y)\right)},
\end{align*}
here 
\begin{align*}
r_{m(n)} = \sqrt{(c^x - (c_m^{\text{AP},x} + nd^{x}_m))^2 + (c^y - (c_m^{\text{AP},y} + n d_m^{y}))^2},
\end{align*}
for $n\in \{-N,...,N\}$. When the evaluation point coincides with the actual user location, i.e., $c^x = c_k^x$ and $c^y = c_k^y$, since the signal subspace and the noise subspace are orthogonal, the following orthogonality condition holds:
\begin{align*}
\tilde{\pmb{v}}^H(c^x, c^y) \, \pmb{U}_{\text{n}} \, \pmb{U}_{\text{n}}^H \, \tilde{\pmb{v}}(c^x, c^y) = 0.
\end{align*}
Consequently, the positions of all $K$ users are estimated as the $K$ dominant peaks of the MUSIC pseudo-spectrum:
\begin{align*}
P_{\text{MUSIC}}(c^x, c^y)
= \frac{1}{\tilde{\pmb{v}}^H(c^x, c^y) \, \pmb{U}_{\text{n}} \, \pmb{U}_{\text{n}}^H \, \tilde{\pmb{v}}(c^x, c^y)},
\end{align*}
where the estimated locations $(\hat{c}_k^x, \hat{c}_k^y)$ correspond to the $K$ largest peaks of $P_{\text{MUSIC}}(c^x, c^y)$.
\ifCLASSOPTIONcaptionsoff
  \newpage
\fi


 \vspace{-0.2cm}
\bibliography{bare_jrnl}
\bibliographystyle{IEEEtran}
%

%




\end{document}